\documentclass[prd,aps,twocolumn,nofootinbib,showpacs,floats,letterpaper,floatfix,groupedaddress,eqsecnum]{revtex4}

\usepackage{dcolumn,epsfig}
\usepackage{amssymb,amsmath}

\def\nn{\nonumber}

\def\pa{\partial}

\def\be{\begin{equation}}
\def\ee{\end{equation}}
\def\beq{\begin{eqnarray}}
\def\eeq{\end{eqnarray}}

\def\IL{\relax{\rm I\kern-.18em L}}

\def\nn{\nonumber}

\begin{document}

\title{Are black holes in alternative theories serious astrophysical candidates? \\The case for Einstein-Dilaton-Gauss-Bonnet black holes}



\author{Paolo Pani} \email{paolo.pani@ca.infn.it}
\affiliation{Dipartimento di Fisica, Universit\`a di Cagliari, and INFN sezione di Cagliari, Cittadella Universitaria 09042 Monserrato, Italy \\Currently at Centro Multidisciplinar de Astrof\'{\i}sica - CENTRA,
Dept. de F\'{\i}sica, Instituto Superior T\'ecnico, Av. Rovisco
Pais 1, 1049-001 Lisboa, Portugal }

\author{Vitor Cardoso} \email{vcardoso@fisica.ist.utl.pt}
\affiliation{Centro Multidisciplinar de Astrof\'{\i}sica - CENTRA,
Dept. de F\'{\i}sica, Instituto Superior T\'ecnico, Av. Rovisco
Pais 1, 1049-001 Lisboa, Portugal \& \\  Department of
Physics and Astronomy, The University of Mississippi, University,
MS 38677-1848, USA}

\begin{abstract}

It is generally accepted that Einstein's theory will get some as yet unknown corrections, possibly large in the strong field regime.
An ideal place to look for these modifications is around the vicinities of compact objects such as black holes.
Our case study here are Dilatonic Black Holes, which arise in the framework of Gauss-Bonnet couplings and one-loop corrected four-dimensional effective theory of heterotic superstrings at low energies. These are interesting objects as a prototype for alternative, yet well-behaved gravity theories: they evade the ``no-hair'' theorem of General Relativity but were proved to be stable against radial perturbations.

We investigate the viability of these black holes as astrophysical objects and try to provide some means to distinguish them from black holes in General Relativity. We start by extending previous works and establishing the stability of these black holes against axial perturbations. We then look for solutions of the field equations describing slowly rotating black holes and study geodesic motion around this geometry. Depending on the values of mass, dilaton charge and angular momentum of the solution, 
one can have differences in the ISCO location and orbital frequency, relatively to black holes in General Relativity. In the most favorable cases the difference amount to a few percent. Given the current state-of-the-art, we discuss the difficulty to distinguish the correct theory of gravity from EM observations or even with gravitational wave detectors.

\end{abstract}


\maketitle

\section{Introduction}

Einstein's theory of General Relativity (GR) has passed numerous consistency and experimental tests in a spectacular way \cite{Will:2005va}. 
Nevertheless, it is a general consensus that GR will get modified at some scale, if only because GR has resisted all attempts at its quantization. 
Moreover, other theories of gravity also pass the experimental tests, some with a better ``quantum behavior''
and should be taken as serious candidates. Unfortunately, the majority of these alternative theories are vastly more complex than GR in their full-fledged form. It is thus not surprising that progress in understanding the exact differences between one and the other and specially differences one can measure experimentally have been slow and mostly focusing on the weak-, far-field behavior.

One of the candidates for a theory of quantum gravity is string theory \cite{polchinski}. Since it is
still difficult to study geometrical settings in superstring theories, most analyses have been performed by using low-energy string-inspired effective theories \cite{Callan:1986jb}. 
Thus, quantum gravity predictions can be tested by studying modifications of GR due to some low-energy truncation of string theory \cite{Gross:1986iv}. Typically the effective theories are supergravities involving not only the metric but also a scalar field (the dilaton) as well as several gauge fields \cite{Zwiebach:1985uq}. 
One of such theories is the one-loop corrected four-dimensional effective theory of the heterotic superstrings
at low energies and a simple particular case is known as Einstein-Dilatonic-Gauss-Bonnet (EDGB) theory (see for instance Ref. \cite{Moura:2006pz} and references cited therein for a nice and concise introduction to this theory). 
In EDGB theory the gauge fields are neglected and only the (spacetime-dependent) coupling between the dilaton and the gravity is considered, with the anomaly-canceling Chern-Simons term also neglected (see for instance \cite{Alexander:2007kv,Smith:2007jm,Alexander:2007zg,Alexander:2007vt} 
for work taking this term into account). At the first order in the Regge slope, $\alpha'$, higher-derivative gravitational terms such as the Gauss-Bonnet (GB) curvature-squared term are present in the action, hence the name. The GB terms avoid some pathological features, for example the theory is ghost-free. Since the equations of motions are still of second order, EDGB theory provides one of the simplest consistent high-energy modifications to GR. Even though it does not seem to be a viable cosmological model \cite{EspositoFarese:2003ze}, the Parametrized Post-Newtonian \cite{Will} expansion of this theory is identical (to lowest order) to that of GR \cite{EspositoFarese:2004cc,Sotiriou:2006pq}, which means that it passes all Solar system-like experimental tests of gravity.  Differences arise only from full nonlinear effects and the ideal place to look for these is near compact objects such as black holes. 

Dilatonic Black Holes (DBHs) do exist in EDGB theory \cite{kanti,Torii:1996yi,Alexeev:1996vs}. They have a regular event
horizon and the geometry is asymptotically flat at infinity (recently DBHs in higher dimensions \cite{Guo:2008hf} and with non-flat asymptotical geometries \cite{Guo:2008eq} have also been studied). In what follows we refer to DBHs in EDGB theory only. DBHs are interesting for many theoretical and practical reasons. DBHs can evade the classical ``no-scalar-hair" theorem~\cite{bek}, and be dressed with classical non-trivial dilaton hair. This is a direct consequence of the GB term~\cite{kanti} and opens up the exciting possibility of ruling out such objects and theories by testing the no-hair theorem, either by gravitational-wave observations of ringdown \cite{Berti:2005ys} or by observations of highly eccentric orbits around supermassive black holes \cite{Will:2007pp}.
The lack of experimental data and some problems on the theoretical side make it very difficult to explore string theory and other theories of quantum gravity. Since quantum gravity effects are expected to play a significant role in strong gravity regime such as in cosmology \cite{Bamba:2007ef} and in black hole physics, the investigation of BHs in EDGB theory can shed new light on some aspects of quantum gravity and/or be used to develop testable predictions of the theory. 

\subsection{Astrophysical implications}

Our purpose here is to begin exploring differences between GR and alternative theories that could be experimentally tested through astrophysical observations, in the strong field regime. 
Thus, even though the theory with which we will work is only a first order truncation of the full action, we will elevate 
it to the status of a complete theory, therefore we place no restriction on the Regge slope.
The first important problem concerns the stability of DBHs. Credible alternatives to the Schwarzschild and Kerr metric of GR must be stable spacetimes. It has been shown \cite{kanti-lin,torii} that DBHs are stable against a small subset of all possible perturbations, linear \emph{radial} perturbations. In this work we characterize completely half of the possible degrees of freedom, by studying general \emph{axial} perturbations.
We find that DBHs are stable also against these perturbations. By itself this is an interesting result confirming that high-energy contributions lead to viable alternatives to classical BHs arising in GR. The viability of DBHs poses the following question: can one devise observational tests to discern a DBH from a classical BH? 
In classical Einstein-Maxwell theory BHs are characterized by three parameters \cite{Hawking:1971vc}: mass $M$, electric charge $Q$ and angular momentum $J \equiv a M \leqslant M^2$. Astrophysical BHs are likely to be electrically neutral because of the effect of surrounding plasma
\cite{Blandford:1977ds}, and therefore tests of alternative theories of gravity can in general focus on rotation alone. The task is still highly non-trivial and as we mentioned earlier, strong-field effects must be searched for in theories that are not already ruled out by Solar-system experimental data \cite{Sotiriou:2006pq}.

Most if not all of present-day astrophysical observations related to compact objects, concern directly or indirectly the motion of matter.
Thus, a study of geodesic motion around compact objects in alternative theories is of utmost importance. 
Geodesics convey very important information on the background geometry. In particular circular orbits whose radius is close to the horizon may be extremely useful, because they already probe strong field regions. They can be used to compute the ``innermost-stable-circular-orbit'' (ISCO), a notion which is very important for interpretation of the experimental data concerning astrophysical black holes. For the Schwarzschild spacetime, $r_{\text{ISCO}}=6M$, while for an extremal rotating Kerr geometry, $r_{\text{ISCO}}=M$,~$9M$ for co- and counter-rotating circular orbits respectively. Measurements of the ISCO are also useful to evaluate the angular momentum of Kerr BHs \cite{Narayan:2005ie}. Current methods to measure the ISCO include spectral fitting, quasi-periodic oscillations and relativistic iron line measurements \cite{Narayan:2005ie}.
Here we show that differences in the ISCO of slowly rotating DBHs and classical Kerr black holes can be significant, depending on the coupling parameter. These differences, which may be detectable in near-future experiments, are likely to increase for highly spinning black holes (which are unfortunately out of the scope of this work). In fact, it seems an exciting possibility that the GRAVITY experiment \cite{Eisenhauer:2008tg}, designed to make precision measurements of orbits of stars in the neighborhood of the black-hole in the center of our galaxy, might already be able to discriminate between these black hole solutions and GR solutions, or otherwise impose stringent bounds on the coupling parameter.

Current techniques to evaluate the angular momentum of astrophysical compact objects are based only on ISCO measurements \cite{Narayan:2005ie}. One needs to \emph{assume} a particular theory of gravity in order to evaluate $J$. Thus discerning a DBH from a Kerr black hole by ISCO measurements is not an easy task. Future gravitational wave experiments will provide a viable method to measure $M$ and $J$ independently \cite{Berti:2004bd}. The analysis below suggests that a \emph{possible} deviation from the expected ISCO in GR can be explained in term of dilatonic charged BH. Therefore in a near future, gravitational wave astronomy may offer a the possiblity to explore string theory-inspired modifications of GR.

Finally, null unstable geodesics are closely related to the appearance of compact objects to external observers \cite{podurets,amesthorne} and have been associated with the characteristic, or quasinormal modes (QNMs) of BHs \cite{Cardoso:2008bp,cardosotopical}. Quasinormal modes are very important in devising experimental tests of GR and for gravitational wave astronomy. Measuring QNM frequencies may provide a definitive proof of the existence of BHs in GR and it could be useful to study corrections to GR too. Thus the analysis of geodesic motion around a DBH can shed new light on various and important aspects of high-curvature corrections to gravity. 

Other theories might also suggest a different ISCO location, different quasinormal modes, etc. Thus, a deviation in these quantities is not a verification of a particular theory, but is a first step in understanding what different theories and scenarios predict in the strong field regime, which could potentially discriminate GR from other alternatives.

The paper is organized as follows. In Section \ref{formalism} we briefly review the main aspects of BHs in EDGB theory. In Section \ref{stability} we prove the linear stability of DBHs against axial perturbations. We proceed by studying slowly rotating DBHs in Section \ref{smallrot}. We prove that such slowly rotating solutions do exist and we characterize them, including a discussion on the ergoregion in these spacetimes. Section \ref{geodesics} discusses geodesics in both spherically symmetric and slowly rotating DBHs as well as the possible experimental tests which can be performed to discern a DBH from a Kerr BH. We compute the ISCO dependence on the angular momentum and the QNM frequencies for a spherical symmetric DBH in the eikonal limit. Conclusions are discussed in Section \ref{conclusion}. In our notation, we use the signature $(+---)$ and the curvature tensor is defined by $R^a_{ijk}=\pa_j\Gamma^a_{ik}+...$.

\section{Spherically symmetric BHs in Einstein-Dilaton-Gauss-Bonnet theory}\label{formalism}
We consider the following low-energy effective action for the heterotic string  \cite{Zwiebach:1985uq}
\be
S=\frac{1}{2}\int d^4 x \sqrt{-g} \left (R-\frac{1}{2}\pa_\mu \phi\pa^\mu\phi+\frac{\alpha' e^\phi}{4 g^2} {\cal R}^2_{GB} \right)\,,\label{eq:action}
\ee
where
\begin{equation}
{\cal R}^2_{GB}=R_{\mu\nu\rho\sigma}R^{\mu\nu\rho\sigma}-
4R_{\mu\nu}R^{\mu\nu}+R^2\,,
\end{equation}
is the GB invariant, $\alpha'$ is the Regge slope and $g^2$ is some gauge coupling constant. We set $g=1$ for the rest of this paper. String-inspired ${\cal O}(\alpha')$ corrections to Einstein's gravitation are included in the action (\ref{eq:action}), while gauge fields and matter are omitted for simplicity. We also note that there is some arbitrariness in the coupling, depending on which frame we take as fundamental \cite{Guo:2008hf}, we choose to keep the $e^{\phi}$ coupling in line with previous works.
We shall refer to the action (\ref{eq:action}) as EDGB theory.

The dilaton field and Einstein's equations derived from (\ref{eq:action}) are
\beq
\frac{1}{\sqrt{-g}} \partial _\mu [\sqrt{-g} \partial ^\mu \phi ]
&=&\frac{\alpha'}{4} e^\phi {\cal R}^2_{GB}\,,
\label{eq:dilaton} \\
G_{\mu\nu} =\frac{1}{2} \partial _\mu \phi
\partial _\nu \phi &-&\frac{1}{4} g_{\mu\nu} (\partial _\rho \phi )^2  -
\alpha' {\cal K}_{\mu\nu}\,,
\label{eq:eins}
\eeq
where $G_{\mu\nu}=R_{\mu\nu} - \frac{1}{2} g_{\mu\nu} R$ is the Einstein tensor and
\beq
{\cal K}_{\mu\nu}=(g_{\mu\rho}g_{\nu\lambda}+g_{\mu\lambda}g_{\nu\rho})
\eta^{\kappa\lambda\alpha'\beta} \nabla _g
[{\tilde R}^{\rho g}_{\,\,\,\,\,\alpha'\beta} \partial _\kappa f]\,.\label{eq:kappamunu}
\eeq
Here,
\beq
&~&\eta ^{\mu\nu\rho\sigma} = \epsilon ^{\mu\nu\rho\sigma}
(-g)^{-\frac{1}{2}}\,,\qquad \epsilon ^{0ijk} = -\epsilon_{ijk} \nn\,,\\
&~&{\tilde R}^{\mu\nu}_{\,\,\,\,\,\kappa\lambda} = \eta^{\mu\nu\rho\sigma}
R_{\rho\sigma\kappa\lambda}\,,\qquad f =  \frac {e^\phi} {8}\nn\,.
\eeq
From the right-hand-side of the modified Einstein's equation (\ref{eq:eins}), one can construct a conserved ``energy momentum tensor'', $\nabla _\mu T^{\mu\nu} = 0$,
\be
T_{\mu\nu} =  -\frac{1}{2} \partial_\mu\phi \partial_\nu \phi + \frac{1}{4}
g_{\mu\nu} (\partial _\rho \phi )^2  + \alpha' {\cal K}_{\mu\nu}\,.
\label{eq:EStens}
\ee
In Ref.\cite{kanti} is shown that the time component of $-T_{\mu\nu}$, which in Einstein's gravity would correspond to the local energy density ${\cal E}$, {\it can be negative}. The reason is that, as a result of the higher derivative GB terms, there are contributions of the gravitational field itself to $T_{\mu\nu}$. The positiveness of $-T_{00}$ is one of the main assumptions of the no-scalar-hair theorem \cite{bek} which can be (and indeed is) evaded in EDGB theory.

\begin{center}
\begin{figure}[ht]
\begin{tabular}{c}
\epsfig{file=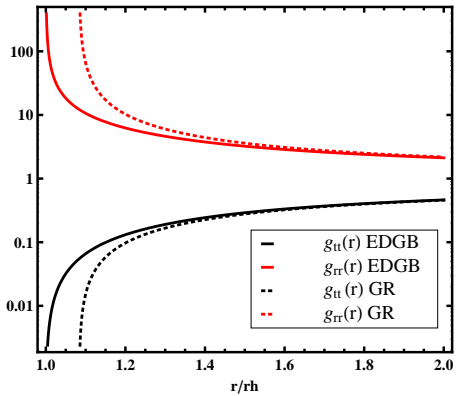,width=180pt,angle=0}
\end{tabular}
\caption{Metric components $g_{tt}$ and $g_{rr}$ for a DBH (solid lines) compared with an equal mass Schwarzschild hole (dotted lines) ($\alpha'/M^2\sim0.691$, which corresponds to $D/M\sim0.572$).}
\label{fig:backfield}
\end{figure}
\end{center}

We now focus on BH solutions in EDGB theory, considering the following spherically symmetric ansatz for the metric 
\be
ds^2=e^{\Gamma(r,t)} dt^2-e^{\Lambda(r,t)} dr^2 -r^2(d\theta^2
+\sin ^2\theta \, d\varphi^2)\,.
\label{2}
\ee
The equations of motion derived from (\ref{eq:dilaton}) and from (\ref{eq:eins}) can be found in Ref. \cite{kanti-lin}.
In a static, asymptotically flat geometry, black hole solutions exist only if \cite{kanti}
\begin{equation}
e^{\phi_h} \le \frac{r_h^2}{\alpha'\sqrt{6}}\,,
\label{eq:cond}
\end{equation}
where $r_h$ and $\phi_h$ are quantities evaluated on the horizon. In particular black hole solutions may exist only for $\alpha'>0$. 
From the asymptotic behavior of the fields one can extract the ADM mass, $M$ and the charge $D$. As shown in Ref.\ \cite{kanti} $M$ and $D$ are not independent quantities, thereby leading to the secondary nature of the dilaton hair~\cite{wilczek}.
These black hole solutions are uniquely characterized by two parameters $(\phi_h, r_h)$, which correspond to a unique choice of $(M,D)$. The equations of motion remain invariant under
a shift $\phi~\rightarrow~\phi+\phi_0$ and a simultaneous radial rescaling $r \rightarrow r e^{\phi_0/2}$. As a consequence of the radial rescaling, the two other asymptotic parameters, $M$ and $D$, are also rescaled according to the rule $M \rightarrow M e^{\phi_0/2}$ and $D \rightarrow D e^{\phi_0/2}$. Due to the above
invariance it is sufficient to vary only one of $r_h$ and $\phi_h$.
Following \cite{kanti-lin} we choose to keep $r_h$ fixed and to vary $\phi_h$. Typical background fields are shown in Fig. \ref{fig:backfield}. Differences in the metric coefficients occur only very close to the horizon. We checked our numerical solutions reproducing results shown in the Table 1 of Ref.\ \cite{kanti}. 

\begin{center}
\begin{figure}[ht]
\begin{tabular}{c}
\epsfig{file=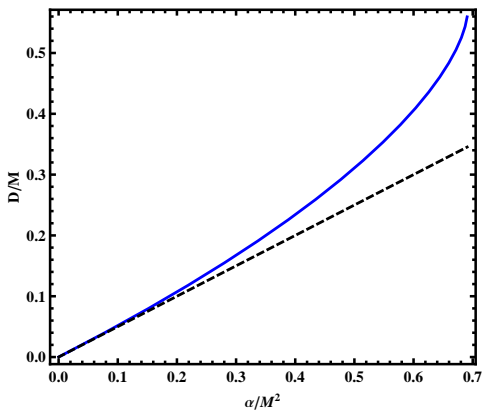,width=180pt,angle=0}
\end{tabular}
\caption{The dilaton charge, $D$, as a function of (re-scaled) $\alpha'/M^2$ (solid line). The whole range of $\alpha'/M^2$ in which a black hole solution exists is shown. Numerical solution behaves as $D/M\sim\alpha'/(2M^2)$ in the $\alpha'\rightarrow0$ limit \cite{Mignemi:1992nt} (dashed line).}
\label{fig:Dcharge}
\end{figure}
\end{center}

After the rescaling, equation (\ref{eq:cond}) can be written in terms of the coupling constant
\be 
0<\frac{\alpha'}{M^2}\lesssim0.691\label{eq:range}\,.
\ee
The maximum value $\alpha'/M^2\sim0.691$ corresponds to $D/M\sim0.572$. For larger values of the coupling constant no DBH solution exists. The dilaton charge, $D$, as a function of $\alpha'/M^2$ is shown in Figs. \ref{fig:Dcharge}, where it is compared with the analytical solution in the $\alpha'\rightarrow0$ limit, $D/M=\alpha'/(2M^2)$ \cite{Mignemi:1992nt}. 
%
\section{\label{stability} Linear stability analysis}
The framework for a linear stability analysis of black holes in theories with Gauss-Bonnet terms was laid down
by Dotti and Gleiser \cite{Dotti:2004sh,Dotti:2005sq,Gleiser:2005ra} in higher dimensions (with no dilaton), and generalized by Moura and Schiappa \cite{Moura:2006pz} in the context of Riemann tensor, ${\cal R}^2$, corrections. Perturbations of four-dimensional DBHs were considered by Kanti {\it et al} \cite{kanti-lin}
and Torii and K.~i.~Maeda \cite{torii}. Unfortunately these authors considered only a very specific type of perturbations, here we want to generalize their results. Consider therefore perturbing the spacetime in the following linear way
\beq
g_{\mu\nu}(x^\rho)&=& g^{(0)}_{\mu\nu}(x^\rho)+\epsilon\, h_{\mu\nu}(x^\rho)\,,\nonumber\\
\phi(x^\rho)&=&\phi_{0}(x^\rho)+\epsilon\, \delta\phi(x^\rho)\,,\label{eq:metricexp}
\eeq
where $\epsilon\ll 1$, $g^{(0)}_{\mu\nu}$ and $\phi_0(x^\rho)$ are the background fields, while $h_{\mu\nu}$ and $\delta\phi(x^\rho)$ are
the perturbations. The background metric $g^{(0)}_{\mu\nu}$ and dilaton field $\phi_0(x^\rho)$ are given by the numerical static black hole solution
described above \cite{kanti}. 

\subsection{General formalism}
To study this problem we use the approach first described by Regge and Wheeler \cite{regge}. After a decomposition in tensorial spherical harmonics \cite{zerilli,mathews}, the perturbations fall into two distinct classes: \emph{axial} (odd) with parity $(-1)^{l+1}$ and \emph{polar} (even) with parity $(-1)^l$, where $l$ is the angular momentum of the particular mode.
The theory described by (\ref{eq:action}) is invariant under diffeomorphisms\footnote{Despite the covariant derivative in eq. (\ref{eq:kappamunu}), equations of motion \emph{do not} contain higher derivatives of the metric $g_{\mu\nu}$ because of the GB term (see equation (4) in \cite{torii}). This allows one to use the same gauge transformations first proposed in Ref.\ \cite{regge}.}, as Einstein's theory is. We can then use the gauge freedom in order to simplify the elements $h_{\mu\nu}$. In the classical Regge-Wheeler gauge the canonical form for the metric perturbations
is (see also Ref.\ \cite{vish}):\\
\noindent {- axial perturbations:}
\begin{eqnarray}
h_{\mu \nu}= \left[
 \begin{array}{cccc} 
 0 & 0 &0 & h_0 
\\ 0 & 0 &0 & h_1
\\ 0 & 0 &0 & 0
\\ h_0 & h_1 &0 &0
\end{array}\right] e^{-i \omega t}
\left(\sin\theta\partial_\theta\right)
P_l\,,
\label{eq:axial}
\end{eqnarray}
{- polar perturbations:}
\begin{eqnarray}
h_{\mu \nu}= \left[
 \begin{array}{cccc} 
 H_0 e^{\Gamma(r)} & H_1 &0 & 0 
\\ H_1 & H_2e^{\Lambda(r)}  &0 & 0
\\ 0 & 0 &r^2K & 0
\\ 0 & 0 &0 & r^2K\sin^2\theta
\end{array}\right] e^{-i \omega t}
P_l.
\nonumber
\end{eqnarray}
Where $P_l=P_l(\cos\theta)$ is the Legendre polynomial with angular
momentum $l$ and $h_0$, $h_1$, $H_1$, $H_2$ and $K$ are unknown radial functions.

Perturbations of the dilaton field, $\delta\phi$, do not appear in the axial equations (see also \cite{holzhey}). The linear stability analysis proceeds by mapping the system of the equations of motion for the perturbation fields under consideration to a stationary one-dimensional Schr\"odinger problem, in an appropriate potential well, in which the `squared frequencies' $\omega ^2$ are the ``energy eigenvalues''. 
Instabilities, then, correspond to bound states, i.e. to negative energy eigenstates or equivalently to frequencies $\omega$ with a positive imaginary component.

Presumably because polar perturbations are extremely complex to analyse, Kanti et al. \cite{kanti-lin} focused on a certain subset, the radial perturbations. They found that the spacetime was stable against radial perturbations, their results being confirmed in Ref. \cite{torii}.
Thus, there are good indications that the spacetime is stable under polar perturbations.
We thus focus here on the other set of perturbations, axial perturbations, which as far as we know are not dealt with in the literature.

\subsection{Axial perturbations}

For axial perturbations only 3 nontrivial Einstein equations can be obtained by substituting (\ref{eq:axial}) and (\ref{eq:metricexp}) into (\ref{eq:eins}). The zeroth-order equations are identically zero, due to the background solution, while the equations for the perturbations, $h_0$ and $h_1$, read:
\beq
{(\varphi,\theta) :\,\,\,\,\, } &&h_0(r)+A_1\, h_1(r)+A_2\, h_1'(r)=0\,,\label{eq:32}\\
{(\varphi,r) :\,\,\,\,\, } &&h_0'(r)-\frac{2}{r} \,h_0(r)+{1,l}\, h_1(r)=0\,,\label{eq:31}\\
{(\varphi,t) :\,\,\,\,\, } &&C_{1, l} \,h_0(r)+C_{2, l}\, h_1(r)+C_{3, l}\, h_0'(r)+\nn\\
&&C_{4, l}\, h_1'(r)+C_{5, l}\, h_0''(r)=0\,,\label{eq:30}
\eeq
where $A_i$, $B_{i, l}$ and $C_{i, l}$ depend on the radial background function $\Gamma, \Gamma', \Gamma'', \Lambda,  \Lambda', \phi_0, \phi_0'$ and $\phi_0''$ found in \cite{kanti}. Their explicit form can be found in Appendix \ref{app:linstab}. We observed numerically that equation (\ref{eq:30}) is automatically satisfied as a consequence of the other two equations and of background solutions. So we are left with a system of two ODEs for two unknown functions $h_0(r)$ and $h_1(r)$. Eliminating $h_0$ from the first order equation (\ref{eq:32}) we obtain a second order differential equation for $h_1$ which can be recast (see Appendix \ref{app:linstab} for details) in the following Schr\"odinger-like equation,
\be
u''(r)+\left[V(r)+\omega^2 K(r)\right]\,u(r)=0\,,\label{eq:oddfinradial}
\ee
These functions are shown in Fig. \ref{fig:oddpotrad}.

The asymptotic behavior of equation (\ref{eq:oddfinradial}) (see Appendix \ref{app:linstab} for details) is
\beq
u''(r)+\left[\frac{V_h+\omega^2 K_h}{(r-r_h)^2}\right]\,u(r)=0\,,\quad&&r\rightarrow r_h\,,\label{eq:oddhor}\\
u''(r)+\omega^2\,u(r)=0\label{eq:oddinf}\,,\quad&&r\rightarrow \infty\,.
\eeq
The asymptotic solutions are
\beq
u(r) &\sim &  u_0(r-r_h)^{\frac{1}{2}\pm\sqrt{\frac{1}{4}-V_h-\omega^2 K_h}}\,,\quad r\rightarrow r_h\,,\label{eq:BChor1}\\
u(r) &\sim &  u_\infty e^{\pm i\omega r}\,,\quad r\rightarrow \infty\,.\label{eq:BChor2}
\eeq
Figure \ref{fig:vhkh} shows the coefficients $V_h$ and $K_h$ as functions of $\alpha'/M^2$. For a Schwarzschild BH, $V_h\equiv1/4$ and $K_h\equiv4 M^2$. In the EDGB case one finds one finds $V_h\sim1/4$ and $K_h\lesssim 4M^2$ in the whole range (\ref{eq:range}). Thus the asymptotic solution (\ref{eq:BChor1}) simplifies
\be
u(r) \sim  u_0(r-r_h)^{\frac{1}{2}\pm i\,\omega M \sqrt{\frac{K_h}{M^2}}}\,,\quad r\rightarrow r_h\,,\label{eq:BChor1sempl}
\ee
Since $\omega$ is complex, the sign in equations (\ref{eq:BChor2}) and (\ref{eq:BChor1sempl}) above has to be chosen so that the solution is regular on the horizon and at infinity. For unstable modes $\text{Im}(\omega)>0$, thus the choice of the minus and the plus sign in eqs. (\ref{eq:BChor2}) and (\ref{eq:BChor1sempl}) respectively guarantees that the corresponding eigenfunctions will vanish at infinity and at the horizon. We have searched for unstable modes using these boundary conditions. We integrated equation (\ref{eq:oddfinradial}) outward starting from the horizon until we found the eigenfrequency corresponding to a vanish field at infinity. We used a Runge-Kutta 4th order method, considering $r_h=1$ and different $\phi_h$ values corresponding to different BH solutions \cite{kanti}. We systematically span the whole range (\ref{eq:range}) for $l=2$, $3$, $4$. We also randomly span other values for $\alpha'$ and $l$ to no avail: no unstable modes were found. We checked the numerical accuracy of our results by changing numerical parameters such as $r_\infty$. Our results strongly suggest that EDGB black holes are stable against axial perturbations. This completes the previous linear stability analysis \cite{kanti-lin,torii} performed for radial perturbations.
\begin{center}
\begin{figure}[ht]
\begin{tabular}{c}
\epsfig{file=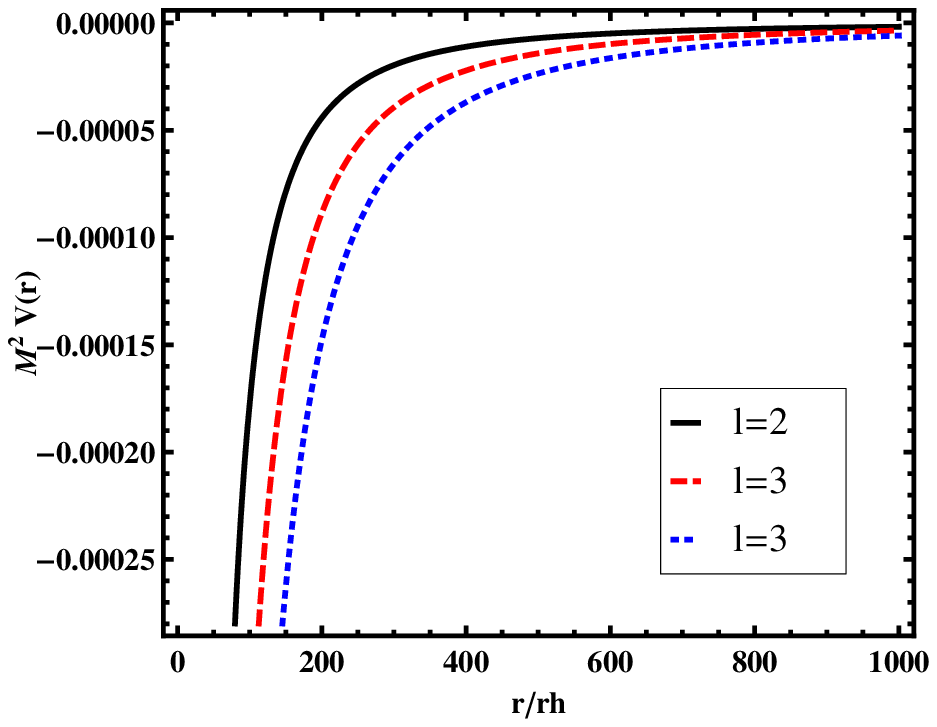,width=180pt,angle=0}\\
\epsfig{file=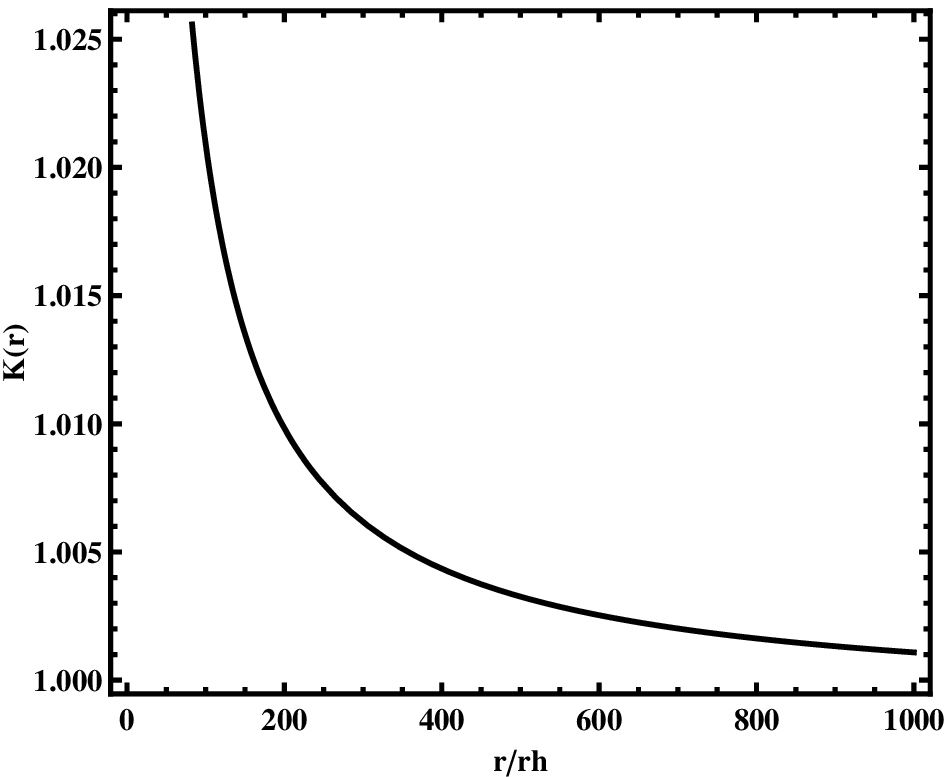,width=180pt,angle=0}
\end{tabular}
\caption{Top Panel: The potential $V(r)$ for the axial perturbations for a EDGB black hole. Different values of the angular momentum $l$ are shown for $\alpha'/M^2\sim0.691$, which corresponds to $D/M\sim0.572$. Bottom Panel: The function $K(r)$ for the axial perturbations equation for a EDGB black hole for $\alpha'/M^2\sim0.691$.}
\label{fig:oddpotrad}
\end{figure}
\end{center}
\begin{center}
\begin{figure}[ht]
\begin{tabular}{c}
\epsfig{file=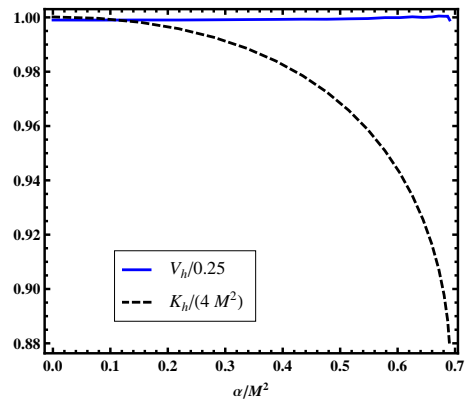,width=180pt,angle=0}
\end{tabular}
\caption{The coefficients $V_h$ (solid line) and $K_h$ (dashed line) as functions of $\alpha'/M^2$ in the whole range (\ref{eq:range}). As for a Schwarzschild BH, $V_h\equiv1/4$, whereas $K_h$ is always smaller than its Schwarzschild counterpart, $K_h\lesssim4 M^2$.}
\label{fig:vhkh}
\end{figure}
\end{center}
It would be of great interest to understand also polar perturbations and reduce them to a single master ordinary differential equation. Due to the complexity of the equations involved we did not perform such an analysis. We note that the present analysis is only possible because axial and polar perturbations decouple in this theory, unlike for instance in Chern-Simons theory \cite{Yunes:2007ss}.

\section{Slowly rotating BHs in EDGB theory}\label{smallrot}

Astrophysical black holes are expected to be highly spinning because of accretion and merger events \cite{Gammie:2003qi,Merritt:2004gc,Berti:2008af,Kesdenreview}. Unfortunately the GB term is very hard to deal with for generic axisymmetric metrics and thus general rotating black holes are difficult to study without a full numerical integration of Einstein's equations. Expansion in $\alpha'$ is another promising approach, which has been successfully implemented for non-rotating DBHs \cite{Mignemi:1992nt}. Here we use another method, searching for solutions describing rotating BHs at every order in $\alpha'$, but in a slow rotation approximation. This method allow us to prove that such (slowly) rotating BHs do exist in EDGB theory. This strongly suggests that, in general, rotating DBHs exist.
In alternative high energy modifications of GR, such as Chern-Simons modified gravity BHs, 
rotating black hole solutions are harder to find \cite{Grumiller:2007rv,Konno:2007ze}.

We follow Hartle's approach \cite{hartle} which is based on axisymmetric perturbations of a spherically symmetric equilibrium solution and on an expansion of the perturbed quantities in a power series of the angular velocity $\Omega$ for frame dragging. The equilibrium solution is the one developed in Ref.\ \cite{kanti} and explored in the previous sections. To first order in $\Omega$ the perturbed metric is
\be
ds^2=e^{\Gamma(r)} dt^2-e^{\Lambda(r)} dr^2 -r^2\left[d\theta^2+\sin ^2\theta \, (d\varphi-\Omega dt)^2\right]
\label{eq:pertmetric}\,,
\ee
where $\Omega\equiv\Omega(r,\theta)$ is the angular velocity $d\phi/dt$ of an observer at $(r,\theta)$ freely falling from infinity. Since we consider only axisymmetric perturbations, we neglected the $\phi$ dependence and we can expand $\Omega(r,\theta)$ in terms of Legendre polynomials
\be
\Omega(r,\theta)=\sum_{l=1}^\infty \Omega_l(r)\left(-\frac{1}{\sin\theta}\frac{dP_l(\cos\theta)}{d\theta}\right)\label{eq:Omega}\,.
\ee
Both the metric and dilaton perturbations are functions of even powers of the angular velocity $\Omega$, thus we can use the unperturbed metric and the unperturbed dilaton field to find $\Omega$-order corrections neglecting $\Omega^2$-order terms. Using the expansion (\ref{eq:Omega}) only the $\{t,\varphi\}$ component of modified Einstein equations (\ref{eq:eins}) is first order in $\Omega$ and, interestingly enough, it gives a separable differential equation as follows
\begin{equation}
\Omega_l''(r)+\frac{G_2(r)}{G_3(r)}\Omega_l'+\left[e^\Lambda \frac{G_1(r)(2-l(l+1))}{G_3(r)}\right]\Omega_l=0\label{eq:eqdiffOmega}\,,
\end{equation}
where
\beq
G_1(r)&=& 2 e^\Lambda+\alpha'e^\phi\left(\Lambda'\phi'-2(\phi'^2+\phi'')\right)\,, \nonumber \\
G_2(r)&=&- e^\Lambda r(-8+r(\Lambda'+\Gamma'))+\nn\\
&-&\alpha' e^\phi\left(\phi'(6-r(3\Lambda'-2\phi'+\Gamma'))+2r\phi''\right)\,, \nonumber \\
G_3(r)&=& 2r^2e^\Lambda -2\alpha're^\phi\phi'\,, \nonumber \\
\eeq
and $\Lambda$, $\phi$, $\Gamma$ are zeroth-order in $\Omega$, therefore given by the spherically symmetric background \cite{kanti}. For $\alpha'=0$ equation (\ref{eq:eqdiffOmega}) is equivalent to the standard GR form (equation  (43) in Ref.\ \cite{hartle} with vanishing stress tensor).

To solve the above equation we must specify the boundary condition for large $r$. Since the scalar field vanishes asymptotically, the solution should approach flat space solution. Thus for large $r$ we can define the angular momentum of the BH via the following behavior,
\be
\Omega_l(r)\sim \frac{2J}{r^3}\,. \label{eq:asymOmega}
\ee
It is worth to note that the dilaton field can introduce order $1/r^3$ corrections to the off-diagonal metric coefficients. In the slowly rotating approach we use, these corrections are neglected. The asymptotic behavior of equation (\ref{eq:eqdiffOmega}) is
\be 
\Omega_l'(r)+\frac{4}{r}\Omega_l'(r)+\frac{2-l(l+1)}{r^2}\Omega_l(r)=0\label{eq:omasymp}\,,
\ee
whose general solution is
\be 
\Omega_l(r)\sim\alpha\, r^{-2-l}+\beta\, r^{-1+l}\label{eq:omsolasymp}\,,
\ee
where $\alpha$ and $\beta$ are constant. With the asymptotic behavior (\ref{eq:asymOmega}), the equation above implies that only the $l=1$ mode is allowed. Thus the equation to solve is
\be
\Omega_1''(r)+\frac{G_2(r)}{G_3(r)}\Omega_1'=0\label{eq:eqdiffOmega1}\,.
\ee
The solution of the above equation is
\begin{equation}
\Omega_1(r)=C_1+C_2\int_{r_h}^r dt{\,e^{-\int_{r_h}^t ds{\,(G_2(s)/G_3(s))}}}\label{eq:solOmega1}\,,
\end{equation}
where the constants $C_1$ and $C_2$ are fixed asking for the asymptotic condition at infinity, eq.~(\ref{eq:asymOmega}). 
The analogous procedure to find slowly rotating Kerr solutions in classical GR is presented in Appendix \ref{ap:slowGR} for completeness,
which is basically a reproduction of the results of Hartle \cite{hartle}.
There we show that this procedure leads to the Kerr metric in lowest order and furthermore that the $l=1$ term is the only possible term in the expansion. Applying the same procedure to a static boson star \cite{bosonstars}, one can prove that such slowly rotating solutions do not exist \cite{Kobayashi:1994qi}.  

\begin{center}
\begin{figure}[ht]
\begin{tabular}{c}
\epsfig{file=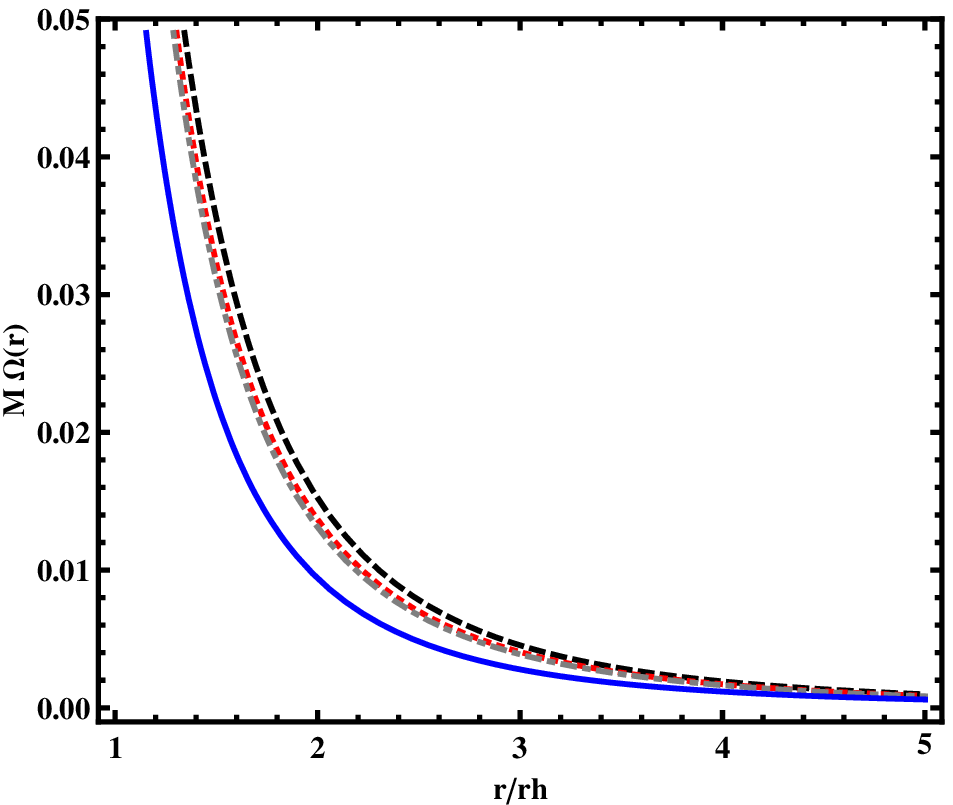,width=180pt,angle=0}\\
\epsfig{file=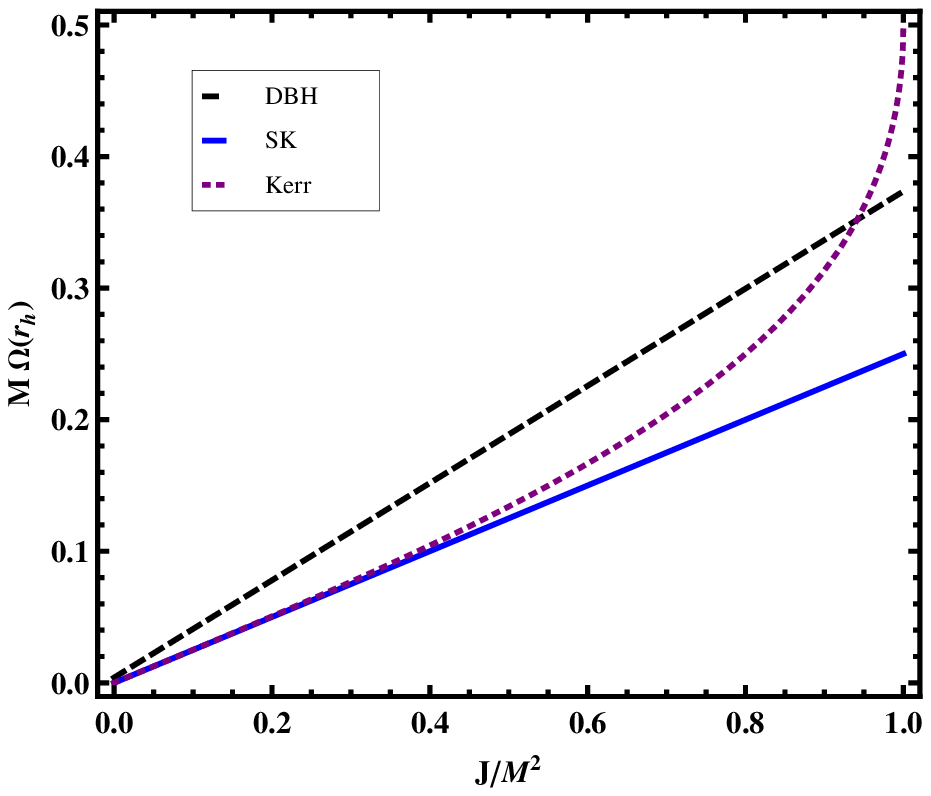,width=180pt,angle=0}
\end{tabular}
\caption{Top Panel: Angular velocity $\Omega$ as a function of $r$ for different slowly rotating BHs in EDGB theory. From top to bottom: decreasing values for $\alpha'/M^2$. For $\alpha'\rightarrow0$ or $D\rightarrow0$ solutions approach the slowly rotating Kerr BH (solid line). Bottom Panel: Angular velocity of the BH, $\Omega_h=\Omega(r_h)$, for a DBH (dashed line) with $\alpha'/M^2\sim0.691$ compared to the same for a slowly rotating Kerr BH (solid line) and a Kerr BH (dotted line). Up to $J/M^2\sim0.5$ results for the slowly rotating Kerr BH reproduce the exact ones. The difference in the angular velocity between slowly rotating DBHs and slowly rotating Kerr BHs and can be as large as $\sim 40\%$.}
\label{fig:omega1}
\end{figure}
\end{center}

In what follows we shall drop the index, $\Omega\equiv\Omega_1$. The angular velocity $\Omega(r)$ for different slowly rotating BHs is shown in Figs. (\ref{fig:omega1}). In the limit $\alpha'\rightarrow0$, one recovers the results for a small rotating BH in General Relativity, obtained perturbing a Schwarzschild BH, as described in the section below. Interestingly, the angular velocity of a BH in the EDGB can be $\sim40\%$ larger than the one for a slowly rotating Kerr BH (see the bottom panel in Fig.(\ref{fig:omega1})) with the same angular momentum. A numerical relation which can be computed from Fig.(\ref{fig:omega1}) is
\be 
M\Omega_h\equiv\ M \Omega(r_h)\sim 0.37\frac{J}{M^2}\,,
\ee
while for a slowly rotating Kerr BH the above relation is $M\Omega_h=0.25\,J/M^2$. Thus, a DBH is more ``compact'' than a BH with same mass and angular momentum. 

%
\subsection{Ergoregion and superradiance}
\begin{center}
\begin{figure}[ht]
\begin{tabular}{c}
\epsfig{file=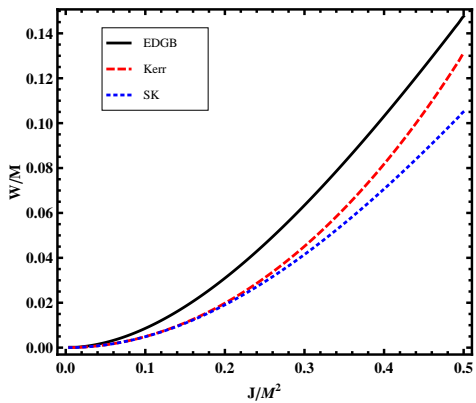,width=180pt,angle=0}
\end{tabular}
\caption{Ergoregion width, $\frac{W}{M}$, for a DBH (solid line), a Kerr BH (dashed line) and a slowly rotating Kerr BH (dotted line) as function of the angular momentum $J$ for $\alpha'/M^2\sim0.691$. Up to $J/M^2\sim0.3$ results for the slowly rotating Kerr BH reproduce the exact ones. The ergoregion width for a rotating DBH can be $\sim50\%$ larger than the one for a Kerr BH.}
\label{fig:ergo}
\end{figure}
\end{center}

Ergoregions can develop in rotating spacetimes. The ergoregion is found by computing the surface on which $g_{tt}$ vanishes. An approximate equation to the location of the ergoregion \cite{schutzexistenceergoregion} is
\be
0=-e^{\Gamma(r)}+\Omega^2(r) r^2\sin^2\theta\,,\label{eqergo}
\ee
which is expected to be a good approximation specially for very compact objects, such as BHs. The solution of
Eq.~(\ref{eqergo}) is topologically a torus. In the equatorial plane we have
\be
r\Omega(r)=\sqrt{e^{\Gamma(r)}}\,.
\label{eqergo2}
\ee
The existence and the boundaries of the ergoregions can be computed from the above equations. The ergoregion width for a DBH compared to both a slowly rotating and a full rotating Kerr BH is shown in Figure \ref{fig:ergo} for different $J$ values. The ergoregion width, $W$, for a Kerr BH on the equatorial plane is $W/M=1-\sqrt{1-(a/M)^2}$. The ergoregion width for a DBH can be $\sim 50\%$ larger than the width for a slowly rotating Kerr BH. Therefore BH superradiance in EDGB is expected to be stronger than in GR. As expected the difference between slowly rotating DBHs and slowly rotating Kerr BHs tends to zero in the $\alpha'\rightarrow0$ limit. 
%
\section{BH Geodesics in EDGB theory}\label{geodesics}

In this section geodesics in the exterior of both non-rotating and slowly rotating EDGB black holes are discussed. 
If we neglect back-reaction effects, which we do in the following, the geodesics in this spacetime correspond to paths followed by time-like or null particles. In principle back-reaction effects should be important for large bodies or strong dilaton fields. However, for small point particles around DBH black holes they should be negligible.
From the analysis of the geodetic motion we compute orbital frequencies related to the ISCO for both non-rotating and slowly rotating DBHs and the QNM frequencies for spherically symmetric BHs in EDGB theory in the eikonal limit. These quantities can be directly measured and can be used as a promising tool to study high-energy modifications of GR.

The motion of test particles around a DBH is deeply related to the coupling between matter and the dilaton. However it turns out that the coupling depends on the choice between two different frames, namely the string frame and the Einstein frame (for a discussion in scalar-tensor theories see Ref. \cite{Faraoni:1999hp} and references therein). The two frames are inequivalent. In the string frame the coupling between matter and the dilaton field is minimal, but Newton constant, $G_N$, depends on the coordinates and light bending is not correctly reproduced \cite{Bruneton:2007si}. Conversely, in the Einstein frame the light bending is correctly reproduced, since $G_N$ is constant, but particles non-minimally couple to the dilaton field. This leads to a violation of the
equivalence principle. This violation is small and compatible with the available tests of the equivalence principle \cite{Faraoni:1999hp} and it is indeed regarded as an important test in discerning the two frames. Which of the two frames is the physically relevant is an open question, which will eventually be settled by experiments.

Following Refs. \cite{kanti} and \cite{Torii:1996yi} we always assume the Einstein frame. In this frame the total action reads
\beq 
S_{\text{TOT}}&&=S+S_{c}+S_{ts}=\label{eq:totaction}\\
&&=S+S_c-m\int dt\,e^{b\phi}\sqrt{-g_{\mu\nu}\frac{\partial x^\mu}{\partial t}\frac{\partial x^\nu}{\partial t}}\,,\nn
\eeq
where $S$ is the EDGB action, eq. (\ref{eq:action}), $m$ is the mass of the test particle and $S_c$ is the action which describes some coupling between the test particle and the dilaton field. The constant $b$ in the equation above is the coupling between the matter and the dilaton field. Its particular value depends on the specific theory from which the low energy theory comes. For low energy modifications from heterotic string theory, $b=1/2$. We shall discuss the motion of a test particle keeping a general value of $b$ and we only specialize to $b=1/2$ when we discussing numerical results. As already mentioned, in the equation above we neglect gauge fields, such as Maxwell fields. Furthermore we neglect any coupling between the dilaton field and the test particle, $S_c$=0, as well as any back-reaction of the background fields.

The string frame and the Einstein frame are related by a Weyl transformation \cite{Casadio:1998wu}. Since null geodesics equations are Weyl invariant, the motion of massless particles is the same in both frames and it is described by the standard geodesics equations.
\subsection{Geodesics: non-rotating case}
Following Chandrasekhar \cite{chandra} and Ref.\ \cite{Cardoso:2008bp} we consider a four-dimensional stationary, spherically symmetric line element
\be
ds^2=f(r)dt^2-\frac{1}{g(r)}dr^2-r^2({d\theta^2+\sin^2\theta}d\varphi^2)\,.\label{eq:metricsphe}
\ee
We also take a spherically symmetric dilaton and set $H(r)=2 b\,\phi(r)$. Due to the symmetry of the above background fields, the trajectory of a particle is planar, say $\theta=\pi/2$ and the following conserved quantities can be defined: the (dimensionless) specific energy of the test particle, $E=e/m$ and the specific angular momentum, $L=\ell/(m)$, where $e$, $\ell$ and $m$ are the energy, the angular momentum and the mass of the test particle respectively. Let us restrict attention to circular orbits, for which the Lagrangian is
\be 2{\cal L}=e^{H(r)}\left(f(r)\,\dot{t}^2-\frac{1}{g(r)}\dot{r}^2-r^2\dot{\varphi}^2\right) \,,
\label{lagrangianads}
\ee
From the Lagrangian above, the equations of motion for the coordinates $x^\mu=(t,r,\theta,\phi)$ read
\beq
{\dot{r}}^2&=&V(r)=\frac{g(r)}{e^{2H(r)}}\left[\frac{E^2}{f(r)}-\frac{L^2}{r^2}-\delta_1\,e^{H(r)}\right]\,,\label{eq:geod}\\
{\dot{\varphi}}&=&\frac{L}{r^2}e^{-H(r)}\,,\qquad{\dot{t}}=\frac{E}{f(r)}e^{-H(r)}\label{eq:geodphit}\,,
\eeq
where $\delta_1=0,1$ for light-like and timelike geodesics respectively and the derivative is intended to respect with the proper time. 
\subsubsection{Time-like geodesics}
We set $\delta_1=1$ in equation (\ref{eq:geod}). The requirement for a circular orbit at $r=r_{t}$ is $V(r_t)=V'(r_t)=0$, thus

\be 
E^2=e^{H_t}\frac{2 f_t^2(1+r_t H'_t/2)}{2 f_t-r_t f_t'}\,,\;\; L^2=e^{H_t}\frac{r_t^3 f_t'(1+H'_tf_t/f'_t)}{2 f_t-r_t f_t'}\,.\nonumber
\ee
We choose the notation $f_t\equiv f(r_t)$, with the subscript $t$ standing for timelike. Since the energy must be real we require
\be 
\frac{2+r_tH'_t}{2 f_t-r_t f_t'}>0\,. \label{eq:geocond1}
\ee
The condition for a stable circular orbit is 
\beq
&&V''_t=2\frac{g_t}{f_t}e^{-H_t}\times\nn\\
&&\left(\frac{(2 (f_t')^2-f_t f_t'')(1+r_t H'_t/2)-3 f_t f'_t/r_t(1+H'_t f_t/f'_t)}{2 f_t-r_t f'_t}+\right.\nn\\
&&\left.-\frac{f_t}{2}[H''_t+(H'_t)^2]\right)<0\,.\label{eq:geostab}
\eeq
From equations (\ref{eq:geodphit}) we define the orbital angular velocity
\be 
\Omega_t=\frac{\dot{\varphi}}{\dot{t}}=\sqrt{\frac{f_t'}{2 r_t}\left(\frac{1+H'_t f_t/f'_t}{1+r_t H'_t/2}\right)}\,.\label{eq:geoOmegaT}
\ee
Equations above reduce to the usual ones for $H(r)\equiv0$.

For the well known Schwarzschild case we have $f(r)=g(r)=1-2M/r$ and $H(r)=0$. In this case the condition (\ref{eq:geocond1}) reads $r_t>3M$ and from condition (\ref{eq:geostab}) for stable orbits, we find $r_t>6M$. For $3M<r_t<6M$ only unstable orbits exist, hence the radius $r=r_{\text{ISCO}}=6M$ is known as ``innermost-stable-circular-orbit'' (ISCO). The orbital angular velocity at the ISCO is $M \Omega_{\rm ISCO}=1/(6\sqrt{6})$. 

\begin{center}
\begin{figure}[ht]
\begin{tabular}{c}
\epsfig{file=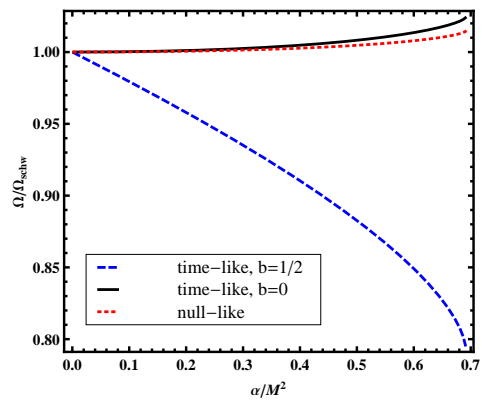,width=180pt,angle=0}
\end{tabular}
\caption{Angular frequency for circular timelike, $\Omega_{\rm ISCO}$, for $b=0$ (solid line) and for $b=1/2$ (dashed line) and for null-like, $\Omega_c$ (dotted line) geodesics, normalized to the Schwarzschild value.}
\label{fig:omega_compar}
\end{figure}
\end{center}

To compute the above quantities in EDGB theory we set $f(r)=e^{\Gamma(r)}$, $g(r)=e^{-\Lambda(r)}$ and $H(r)=\phi(r)$, where $\Gamma(r)$, $\Lambda(r)$ and $\phi(r)$ represent the spherically symmetric BH solution found in Ref.\ \cite{kanti}. We also specialize to the case $b=1/2$. Numerical results for a EDGB non-rotating BH are shown in Table \ref{tab:geod} and Fig. \ref{fig:omega_compar}. The comparison with Schwarzschild BH (with the same mass) and inclusion of null geodesics is discussed in the section below. For completeness we show results both for $b=1/2$ and $b=0$. The effect of a non-vanishing coupling is leading and it makes the angular frequency for a DBH smaller than in the Schwarzschild case (conversely the ISCO is larger). We find that the difference between the orbital frequency in GR and in EDGB theory can be as large as $\sim 20\%$, depending on the coupling constant $\alpha'/M^2$. In the $\alpha'\rightarrow0$ limit Schwarzschild results are recovered. The largest deviations occur when relation (\ref{eq:cond}) is saturated by chosing the appropriate $\phi_h$. This corresponds to the maximum value $\alpha'/M^2\sim0.691$, or equivalently to $D/M\sim0.572$. It is worth noticing that the same occurs for a Reissner-Nordstr\"{o}m BH, for which we have a relation between the electrical charge and the mass, $Q/M<1$. Qualitatively, taking in account the coupling $b$, in EDGB theory the ISCO is always larger than it is for a Schwarzschild BH and the orbital frequency for a timelike geodesic is always smaller. Results for null geodesics do not depend on the coupling $b$, as explained in the section below. In Table \ref{tab:geod} we also show the quantity
\be 
\eta=\frac{E_\infty-E_{\text{ISCO}}}{E_\infty}\nn\,,
\ee
which is the binding energy per unit rest-mass at the ISCO. Because the accretion inside the ISCO is assumed
to be in free-fall, this is equal to the integrated luminosity per unit rest-mass accreted, i.e. the radiative
efficiency of accretion. This efficiency ranges from $\sim5.7\%$ for a Schwarzschild black hole hole to $\sim42\%$ for an extremal Kerr hole.

Finally, we searched for static equilibrium solutions (following Maki and Shiraishi \cite{Maki:1992up}), i.e., a point particle at rest at a distance say, $r=r_0$. For this, we set $L=0$ in Eq. \ref{eq:geod} and 
ask for $V(r_0)=V'(r_0)=0$. We find no solution to these equations, meaning such a configuration does not seem to be possible
for this theory.
\begin{widetext}
\begin{center}
\begin{table}
\caption{\label{tab:geod} Some geodesics-related quantities for a spherically symmetric EDGB BH: orbital frequencies for timelike ($\Omega_{\rm ISCO}$), and null ($\Omega_c$) circular orbits and the radiative efficiency, $\eta$. We compare to the Schwarzschild case, for different values of $\alpha'/M^2$ covering whole spectrum $0<\alpha'/M^2\lesssim0.691$. We also give the percentage ($\%$) difference between EDGB and GR and we show results both for the coupling $b=1/2$ and $b=0$. Null-like geodesics do not depend on the coupling $b$.}
\begin{tabular}{|c|c|c|c|c|c|c|}\hline
$\alpha'/M^2$ & $D/M$ & $M\Omega_{\rm ISCO}$ $(b=0)$ $(\pm\%)$ & $M\Omega_{\rm ISCO}$ $(\pm\%)$  & $M\Omega_c$ $(\pm\%)$ & $\eta$ $(b=0)$ $(\pm\%)$ & $\eta$ $(\pm\%)$ \\ 
\hline\hline
$0$ & $0$ & $0.0681\sim\frac{1}{6\sqrt{6}}$ & $0.0681\sim\frac{1}{6\sqrt{6}}$ & $0.1925\sim\frac{1}{3\sqrt{3}}$ & $0.0572\sim\frac{3-2\sqrt{2}}{3}$ & $0.0572\sim\frac{3-2\sqrt{2}}{3}$\\
$3\times10^{-6}$ & $10^{-6}$ & $0.0681$ $(\sim0\%)$ & $0.0681$ $(\sim0\%)$ & $0.1925$ $(\sim0\%)$ & $0.0572$ $(\sim0\%)$ & $0.0572$ $(\sim0\%)$ \\ 
$3\times10^{-3}$ & $2\times10^{-3}$ & $0.0681$ $(\sim0\%)$ & $0.0680$ $(-0.1\%)$ & $0.1925$ $(\sim0\%)$ & $0.0572$ $(\sim0\%)$ & $0.0571$ $(-0.1\%)$\\ 
$0.027$ & $0.013$ & $0.0681$ $(\sim0\%)$ & $0.0677$ $(-0.5\%)$ & $0.1925$  $(\sim0\%)$ & $0.0572$ $(\sim0\%)$ & $0.0567$ $(-0.9\%)$\\ 
$0.238$ & $0.129$ & $0.0681$ $(+0.2\%)$ & $0.0646$ $(-5.0\%)$ & $0.1926$ $(+0.1\%)$ & $0.0572$ $(+0.1\%)$ & $0.0525$ $(-8.3\%)$\\ 
$0.417$ & $0.246$ & $0.0684$ $(+0.5\%)$ & $0.0617$ $(-9.4\%)$ & $0.1930$ $(+0.3\%)$ & $0.0574$ $(+0.3\%)$ & $0.0484$ $(-15.4\%)$\\ 
$0.545$ & $0.350$ & $0.0687$ $(+1.0\%)$ & $0.0591$ $(-13.1\%)$ & $0.1936$ $(+0.6\%)$ & $0.0575$ $(+0.6\%)$ & $0.0449$ $(-21.6\%)$\\ 
$0.630$ & $0.441$ & $0.0691$ $(+1.6\%)$ & $0.0569$ $(-16.3\%)$ & $0.1942$ $(+0.9\%)$ & $0.0577$ $(+0.9\%)$ & $0.0419$ $(-26.7\%)$\\ 
$0.677$ & $0.516$ & $0.0695$ $(+2.1\%)$ & $0.0552$ $(-18.9\%)$ & $0.1948$ $(+1.2\%)$ & $0.0579$ $(+1.2\%)$ & $0.0396$ $(-30.8\%)$\\ 
$0.691$ & $0.572$ & $0.0697$ $(+2.5\%)$ & $0.0539$ $(-20.9\%)$ & $0.1953$ $(+1.5\%)$ & $0.0580$ $(+1.3\%)$ & $0.0378$ $(-33.9\%)$\\ 
\hline\hline
\end{tabular}
\end{table}
\end{center}
\end{widetext}
%
\subsubsection{Null geodesics}
We now consider null circular geodesics, labeled by a ``c'' subscript. It is easy to show that, setting $\delta_1=0$ in equation (\ref{eq:geod}), the dilaton field, $H(r)$, gives no contribution. Requiring $V_c=V'_c=0$ we find
\be
\frac{E}{L}=\pm\sqrt{\frac{f_c}{r_c^2}}\,,\quad 2 f_c=r_c f_c'\,.\label{eq:geocond2}
\ee
The condition for a circular orbit reads
\be 
V''_c=2\frac{L^2g_c e^{-2H_c}}{r_c^4f_c}\left(2f_c-r_cf_c''\right)\,,\label{eq:geostabN}
\ee
whose sign does not depend on $H(r)$. The orbital angular velocity is
\be 
\Omega_c=\frac{1}{b_c}=\frac{\dot{\varphi}}{\dot{t}}=\frac{\sqrt{f_c}}{r_c}\,,\label{eq:geoOmegaN}
\ee
where $b_c=L/E$ is the impact parameter. In the Schwarzschild case the condition (\ref{eq:geocond2}) restricts the null circular orbits to $r_c=3M$. The orbital angular velocity is $M \Omega_c=1/(3\sqrt{3})$. Results for a EDGB spherically symmetric BH are summarized in Table \ref{tab:geod}. In this case there is no effect from the coupling $b$ and the difference between EDGB and GR theory is of the order $1\%$. From (\ref{eq:geostabN}) we find that $V''_c>0$ and therefore only unstable null circular orbits exist. Hence also in EDGB case null geodesics are always unstable aganist small perturbations. 
\subsubsection{QNMs in the large $l$ limit for spherically symmetric EDGB BHs}
%
\begin{center}
\begin{table}
\caption{\label{tab:QNMs} Real and imaginary part of the QNM frequencies defined by the formula $\omega_{QNM}=\Omega_c\,l-i\,(n+1/2)|\lambda|$ in the eikonal approximation \cite{Cardoso:2008bp}. Terms between parenthesis are the differences with respect to the Schwarzschild case, for which  $M\Omega=M\lambda=1/(3\sqrt{3})$.}
\begin{tabular}{|c|c|c|c|}\hline
$\alpha'/M^2$ & $D/M$ & $M\Omega_c$ $(\pm\%)$  & $M\lambda$ $(\pm\%)$ \\ 
\hline\hline
$3\times10^{-6}$ & $10^{-6}$ & $0.1925$ $(\sim0\%)$ & $0.1925$ $(\sim0\%)$ \\ 
$3\times10^{-3}$ & $2\times10^{-3}$  & $0.1925$ $(\sim0\%)$ & $0.1925$ $(\sim0\%)$ \\ 
$0.027$ & $0.013$ & $0.1925$  $(\sim0\%)$ & $0.1925$ $(\sim0\%)$ \\ 
$0.238$ & $0.129$ & $0.1926$ $(+0.1\%)$ & $0.1924$ $(-0.1\%)$ \\ 
$0.417$ & $0.246$ & $0.1930$ $(+0.3\%)$ & $0.1921$ $(-0.2\%)$ \\ 
$0.545$ & $0.350$ & $0.1936$ $(+0.6\%)$ & $0.1916$ $(-0.4\%)$ \\ 
$0.630$ & $0.441$ & $0.1942$ $(+0.9\%)$ & $0.1909$ $(-0.8\%)$ \\ 
$0.677$ & $0.516$ & $0.1948$ $(+1.2\%)$ & $0.1901$ $(-1.3\%)$ \\ 
$0.691$ & $0.572$ & $0.1953$ $(+1.5\%)$ & $0.1892$ $(-1.7\%)$ \\ 
\hline\hline
\end{tabular}
\end{table}
\end{center}

Particularly promising for gravitational-wave detection are the characteristic vibration modes of black holes \cite{Berti:2005ys,cardosotopical}.
These modes, called quasinormal modes (QNMs), are exponentially damped sinusoids and carry an imprint of the black hole, its features being independent on what exactly excited the modes. QNMs are excited to a large amplitude for instance in the inspiral and subsequent merger of black hole or neutron star binaries. In fact, the ringdown phase of supermassive black holes can have a larger signal-to-noise ratio than any other signal \cite{Flanagan:1997sx,Berti:2005ys,cardosotopical,Berti:2007zu}, justifying dedicated searches for ringdown in current gravitational-wave detectors such as LIGO or TAMA \cite{Goggin:2006bs,Tsunesada:2005fe}.

It was recently shown \cite{Cardoso:2008bp} that there is a simple relation between QNMs and circular null geodesics in general spacetimes in the eikonal, i.e. large-$l$ limit,
\be
\omega_{QNM}=\Omega_c\,l-i\,(n+1/2)|\lambda|\,,\label{eq:QNM}
\ee
where $\Omega_c$ is defined in eq. (\ref{eq:geoOmegaN}) and 
\be 
\lambda=\sqrt{\frac{g_c}{2r_c^2}\left(2f_c-r_c^2f_c''\right)}\,,\label{eq:lyap}
\ee
is the Lyapunov exponent, describing the inverse of the instability timescale of the geodesic. For circular null geodesics the argument of the square root in eq. (\ref{eq:lyap}) is always positive.
QNMs are easly computed from the equation above and again the difference between the real and imaginary part of the QNM frequency for a DBH and for a Schwarzschild BH is order $1\%$, as shown in Table \ref{tab:QNMs}. 
We have not attempted to compute the least damped QNMs for these black holes, but we will assume they behave in the same way as the coupling constant varies \cite{cardosotopical}. Discriminating such small percentage differences is {\it in principle} doable with the gravitational-wave detector LISA \cite{Berti:2005ys,Berti:2007zu}, though it may be very challenging due to systematic errors \cite{Berti:2007fi,Berti:2007dg,cardosotopical,Berti:2005ys,Berti:2007zu}. As far as we know, the QNMs of DBHs have not been considered in the literature, even though their computation is extremely relevant. Nevertheless, the computation of QNMs of black holes in higher-dimensional Gauss-Bonnet theories without a dilaton, or purely dilatonic black holes has been done \cite{Abdalla:2005hu,Konoplya:2004xx,Konoplya:2001ji}, we expect the four-dimensional case to proceed along similar lines.

\subsection{Geodesics: rotating case}
We will now consider geodesics in the the slowly rotating DBHs spacetime studied earlier in Section \ref{smallrot}. For that, let's
take the general four-dimensional static axisymmetric spacetime on the equatorial plane
\be 
ds^2=f(r)dt^2-\frac{1}{g(r)}dr^2-h(r)d\varphi^2+2j(r)dtd\varphi\,,\label{eq:metricRot}
\ee
where we specialize to $\theta=\pi/2$ and $d\theta=0$. We also consider the dilaton field on the equatorial plane and set $H(r)=2 b\phi(r)$. Some interesting special cases of the metric above are listed in Table \ref{tab:spcase}. Following Chandrasekhar \cite{chandra} we consider planar orbits, for which the Lagrangian is
\be
2{\cal{L}}=e^{H(r)}\left(f(r)\dot{t}^2-\frac{1}{g(r)}\dot{r}^2-h(r)\dot{\varphi}^2+2j(r)\dot{t}\dot{\varphi}\right)\,.\label{eq:geodLagr}
\ee
The generalized momenta are
\beq
p_t&=&\text{const}=e^{H(r)}\left[f(r)\dot{t}+j(r)\dot{\varphi}\right]\equiv E\,,\nonumber\\
-p_{\varphi}&=&\text{const}=e^{H(r)}\left[-j(r)\dot{t}+h(r)\dot{\varphi}\right]\equiv L\,,\nonumber\\
-p_r&=&\frac{e^{H(r)}}{g(r)}\dot{r}\,,\nonumber
\eeq
and the Hamiltonian is
\be
2{\cal{H}}=2(p_t\dot{t}+p_{\varphi}\dot{\varphi}+p_r\dot{r}-{\cal{L}})=E\dot{t}-L\dot{\varphi}-\frac{e^{H(r)}}{g(r)}\dot{r}^2=\delta_1\,,\label{eq:geodHam}
\ee
where again $\delta_1=1$, $0$ for time-like and null geodesics respectively. Using the integrals of motion $E$ and $L$ the equations of motion read
\beq
{\dot{r}}^2&=&V(r)=g(r)e^{-2H(r)}\times\label{eq:geodROT}\\
&&\times\left[\frac{h(r)\,E^2-f(r)\,L^2-2j(r)\,E\,L}{{j^2(r)+f(r)h(r)}}-\delta_1e^{H(r)}\right]\,,\nn\\
{\dot{\varphi}}&=&\frac{j(r)\,E+f(r)\,L}{j^2(r)+f(r)h(r)}e^{-H(r)}\,,\nn\\
{\dot{t}}&=&\frac{h(r)\,E-j(r)\,L}{j^2(r)+f(r)h(r)}e^{-H(r)}\nn\,.
\eeq
\begin{center}
\begin{table}
\caption{\label{tab:spcase} {Some particular cases of interest of the metric (\ref{eq:metricRot}) along the equatorial plane: axially and spherically symmetric spacetimes. Then we specialize to the Kerr metric, and its slow rotation limit (Slow Kerr, SK) discussed in Appendix \ref{ap:slowGR}, and finally to the general slowly rotating metric described in Section \ref{smallrot} and then we compare this to the slowly rotating Kerr solution, discussed.}}
\begin{tabular}{|c|c|c|c|c|}\hline
Axial Sym&Sph. Sym&Kerr& SK& Slow rot\\ 
\hline\hline
$f(r)$ & $f(r)$ & $1-\frac{2M}{r}$ & $1-\frac{2M}{r}$ & $f(r)$ \\ 
$g(r)$ & $g(r)$ & $1-\frac{2M}{r}+\frac{a^2}{r^2}$ & $1-\frac{2M}{r}$ & $g(r)$ \\ 
$h(r)$ & $r^2$ & $r^2+a^2+\frac{2a^2M}{r}$ & $r^2$ & $r^2$ \\ 
$j(r)$ & $0$ & $\frac{2aM}{r}$ & $\frac{2J}{r}$ & $r^2\Omega(r)$\\
$H(r)$ & $H(r)$ & $0$ & $0$ & $H(r)$\\
\hline\hline
\end{tabular}
\end{table}
\end{center}
%
\subsubsection{Time-like geodesics}
\begin{center}
\begin{figure}[ht]
\begin{tabular}{c}
\epsfig{file=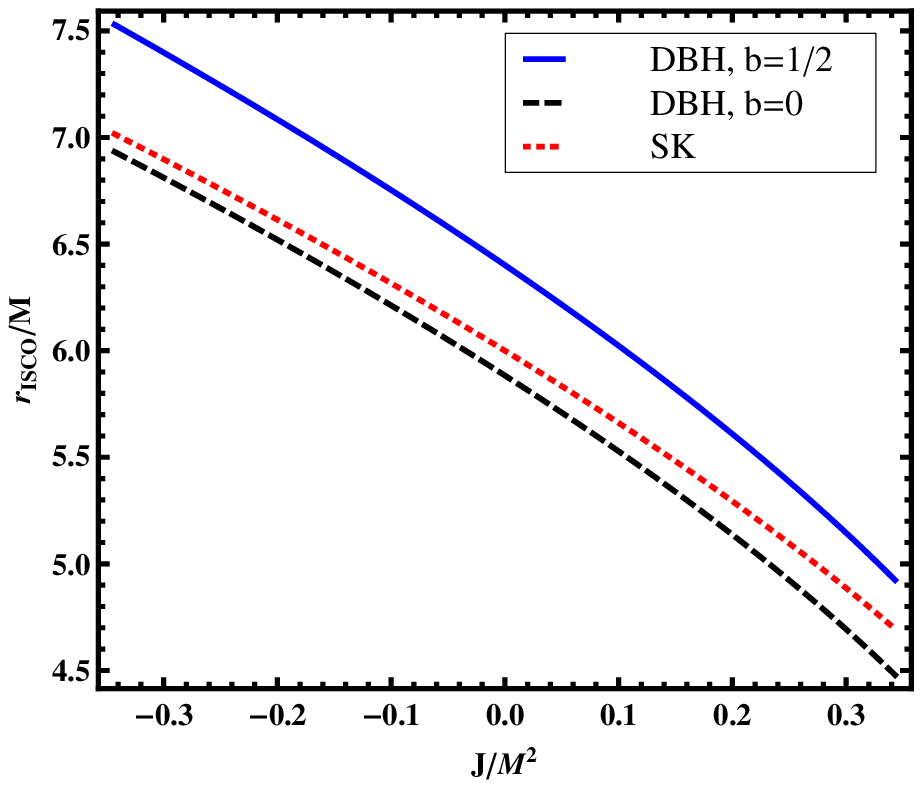,width=180pt,angle=0}\\
\epsfig{file=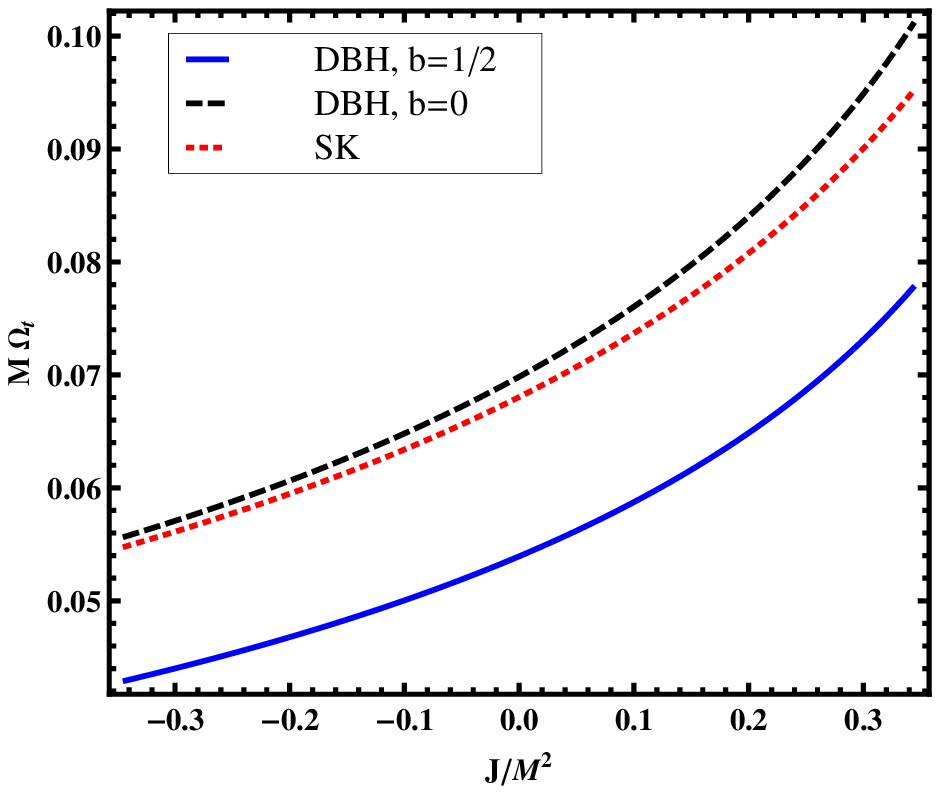,width=180pt,angle=0}
\end{tabular}
\caption{Top Panel: The innermost-stable-circular orbit radius, $r_{\text{ISCO}}/M$ as a function of the angular momentum, $J/M^2$ for a slowly rotating DBH with $\alpha'/M^2\sim0.691$ for $b=1/2$ (solid line), $b=0$ (dashed line) and in the GR limit, $\alpha'/M^2\sim0$ (dotted line). From these plots and from the measurements of $r_{\text{ISCO}}$ the value of the angular momentum $J/M^2$ can be evaluated (see Fig. 2 in Ref. \cite{Narayan:2005ie}). Bottom Panel: the orbital frequency at the ISCO $\Omega_{\rm ISCO}$ as a function of the angular momentum, $J/M^2$ for a slowly rotating DBH with $\alpha'/M^2\sim0.691$ for $b=1/2$ (solid line), $b=0$ (dashed line) and in the GR limit, $\alpha'/M^2\sim0$ (dotted line). Negative values for $J$ correspond to counter-rotating orbits.}
\label{fig:ISCOvsJ}
\end{figure}
\end{center}

Setting $\delta_1=1$ in eq. (\ref{eq:geodROT}) and requiring $V=V'=0$ at the radius $r=r_t$ we find a system of non-linear algebraic equations for $E$ and $L$
\beq
0&=&h_t\,E^2-f_t\,L^2-2j_t\,E\,L-e^{H_t}(j_t^2+f_t\,h_t)\,,\nn\\
0&=&h'_t\,E^2-f'_t\,L^2-2j'_t\,E\,L+\nn\\
&&-e^{H_t}[2\,j_t\,j'_t+f'_t\,h_t+f_t\,h'_t+H'_t(j_t^2+f_t\,h_t)]\,.\nn
\eeq
The system above can be solved analytically but the form of the solutions is not particularly useful. Solutions of the system above can 
be substituted in
\beq 
&V''_t&= \frac{g_t\,e^{-2\,H_t}}{j_t^2+f_t\,h_t}\left[h''_t\,E^2-f''_t\,L^2-2\,j''_t\,E\,L-e^{H_t}\times\right.\nn\\
&\times&\left(\left.[H''_t+(H'_t)^2](j_t^2+f_t\,h_t)+\right.\right.\nn\\
&&\left.\left.+2\,H'_t(2\,j_t\,j'_t+f'_t\,h_t+f_t\,h'_t)+\right.\right.\nn\\
&&\left.\left.+2\,{j'_t}^2+2\,j_t\,j''_t+f''_t\,h_t+2\,f'_t\,h'_t+f_t\,h''_t)\right)\right]\label{eq:VppcSlow}
\eeq
and in 
\be 
\Omega_t=\frac{\dot{\varphi}}{\dot{t}}=\frac{j_t\,E/L+f_t}{h_t\,E/L-j_t}\,.\nn
\ee
From eq. (\ref{eq:VppcSlow}) above we can find the ISCO for a generic rotating BH asking for $V''_t\leq0$. Setting $h(r)\equiv r^2$ and $j(r)\equiv0$ equations above reduce to ones for a spherically symmetric spacetime. The requirement for the energy $E$ to be real imposes
a constraint $r_t>r_{\text{MIN}}$.  Results for circular time-like geodesics are shown in Table \ref{tab:timerotCR} for co-rotating orbits. For $r_{\text{MIN}}<r_t<r_{\text{ISCO}}$ only unstable circular orbits are permitted. For $r_t>r_{\text{ISCO}}$ circular orbits are stable, while for $r_t<r_{\text{MIN}}$ no circular orbits exist. 
\begin{widetext}
\begin{center}
\begin{table}
\caption{\label{tab:timerotCR} {Results for co-rotating time-like (at the ISCO) and null geodesics in slowly rotating EDGB BH spacetimes compared to the slowly rotating Kerr case.}}
\begin{tabular}{|c|c|c|c|c|}\hline
$\alpha'/M^2$ & $J/M^2$ & $M\Omega_{\rm ISCO}$ $(b=0)$ $(\pm\%)$ & $M\Omega_{\rm ISCO}$ $(\pm\%)$  & $M\Omega_c$ $(\pm\%)$ \\ 
\hline\hline
$3\times10^{-3}$ & $0.3$ & $0.0901$ $(\sim0\%)$ & $0.0900$ $(-0.1\%)$ & $0.2220$ $(\sim0\%)$  \\ 
$3\times10^{-3}$ & $0.2$ & $0.0808$ $(\sim0\%)$ & $0.0807$ $(-0.1\%)$ & $0.2101$ $(\sim0\%)$  \\ 
$3\times10^{-3}$ & $0.1$ & $0.0737$ $(\sim0\%)$ & $0.0736$ $(-0.1\%)$ & $0.2005$ $(\sim0\%)$  \\ 
\hline
$0.027$ & $0.3$ & $0.0901$ $(\sim0\%)$ & $0.0896$ $(-0.6\%)$ & $0.2220$ $(\sim0\%)$  \\ 
\hline
$0.238$ & $0.3$ & $0.0903$ $(+0.3\%)$ & $0.0855$ $(-5.1\%)$ & $0.2224$ $(+0.2\%)$  \\ 
\hline
$0.417$ & $0.3$ & $0.0910$ $(+1.0\%)$ & $0.0818$ $(-9.2\%)$ & $0.2232$ $(+0.5\%)$  \\ 
$0.417$ & $0.1$ & $0.0742$ $(+0.6\%)$ & $0.0668$ $(-9.3\%)$ & $0.2012$ $(+0.4\%)$  \\ 
\hline
$0.545$ & $0.3$ & $0.0919$ $(+2.0\%)$ & $0.0787$ $(-12.6\%)$ & $0.2244$ $(+1.1\%)$  \\
$0.545$ & $0.1$ & $0.0746$ $(+1.2\%)$ & $0.0641$ $(-13.0\%)$ & $0.2019$ $(+0.7\%)$  \\  
\hline
$0.630$ & $0.3$ & $0.0929$ $(+3.2\%)$ & $0.0763$ $(-15.3\%)$ & $0.2258$ $(+1.7\%)$  \\ 
$0.630$ & $0.2$ & $0.0827$ $(+2.4\%)$ & $0.0681$ $(-15.8\%)$ & $0.2129$ $(+1.3\%)$  \\ 
$0.630$ & $0.1$ & $0.0751$ $(+1.9\%)$ & $0.0619$ $(-16.1\%)$ & $0.2027$ $(+1.1\%)$  \\ 
\hline
$0.677$ & $0.3$ & $0.0939$ $(+4.3\%)$ & $0.0744$ $(-17.4\%)$ & $0.2272$ $(+2.4\%)$  \\ 
$0.677$ & $0.2$ & $0.0834$ $(+3.2\%)$ & $0.0661$ $(-18.1\%)$ & $0.2139$ $(+1.8\%)$  \\ 
$0.677$ & $0.1$ & $0.0756$ $(+2.5\%)$ & $0.0600$ $(-18.6\%)$ & $0.2035$ $(+1.5\%)$  \\ 
\hline
$0.691$ & $0.3$ & $0.0947$ $(+5.1\%)$ & $0.0729$ $(-19.0\%)$ & $0.2284$ $(+2.9\%)$  \\
$0.691$ & $0.2$ & $0.0839$ $(+3.8\%)$ & $0.0647$ $(-19.9\%)$ & $0.2148$ $(+2.2\%)$  \\
$0.691$ & $0.1$ & $0.0759$ $(+3.0\%)$ & $0.0586$ $(-20.4\%)$ & $0.2041$ $(+1.8\%)$  \\ 
\hline\hline
\end{tabular}
\end{table}
\end{center}
\end{widetext}
Measurements of the ISCO can be used to evaluate the angular momentum of an astrophysical BH \cite{Narayan:2005ie}. Figure \ref{fig:ISCOvsJ} shows how the ISCO and the orbital frequency, $\Omega_t$ depend on $J$ for slowly rotating BHs both in EDGB theory and in GR. Figure \ref{fig:ISCOvsJ} can be used to evaluate the angular momentum once the ISCO has been measured (see Fig. 2 in Ref. \cite{Narayan:2005ie} for details). Again the role of the coupling $b$ is leading for timelike geodesics. As a qualitative result the orbital frequency is smaller for a slowly rotating EDGB BH than the one for a Kerr BH and the ISCO is larger. Differences range from $10\%$ to $\sim 20\%$ depending on the angular momentum $J/M^2$ and on the dilatonic charge $D/M$. The difference is monotonically decreasing for larger rotations, it strongly depends on $b$ but not on $J$. The effect of rotation seems to be subleading.
%
\subsubsection{Null geodesics}
\begin{center}
\begin{figure}[ht]
\begin{tabular}{c}
\epsfig{file=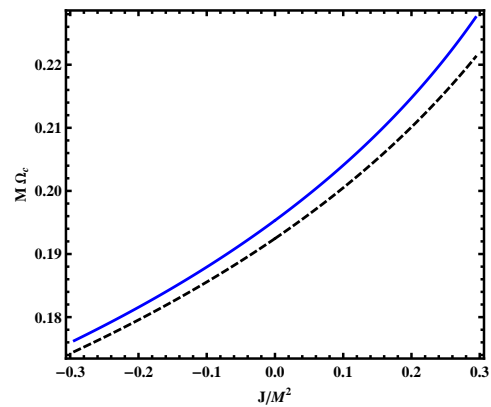,width=180pt,angle=0}
\end{tabular}
\caption{The orbital frequency, $\Omega_c$, for circular null-like orbits as a function of the angular momentum, $J/M^2$ for a slowly rotating DBH with $\alpha'/M^2\sim0.691$ (solid line) and in the GR limit, $\alpha'/M^2\sim0.03$ (dashed line). Negative values for $J$ correspond to counter-rotating orbits.}
\label{fig:OmegaCvsJ}
\end{figure}
\end{center}

Focusing on circular orbits ($\delta_1=0$), we require $V=V'=0$ at the radius $r=r_c$. These two conditions read
\beq
\frac{E}{L}=\frac{j_c}{h_c}\pm\sqrt{\left(\frac{j_c}{h_c}\right)^2+\frac{f_c}{h_c}}\,,\label{eq:slownull1}\\
h'_c\left(\frac{E}{L}\right)^2-f'_c+2\,j'_c\,\frac{E}{L}=0\,.\label{eq:slownull2}
\eeq
In this case
\be
V''_c=\frac{L^2g_ce^{-2H_c}}{j_c^2+f_c\,h_c}\left[h''_c\left(\frac{E}{L}\right)^2-f''_c-2\,j''_c\,\frac{E}{L}\right]\,,\label{eq:Vppcnullrot}
\ee
which is positive at $r=r_c$. As expected the positiveness of equation above does not depend on the dilaton $H(r)$. Therefore only unstable null circular orbits are allowed. The angular velocity is
\be 
\Omega_c=\frac{1}{b_c}=\frac{j_c}{h_c}\pm\frac{\sqrt{j_c^2+f_c\,h_c}}{h_c}\,,\label{eq:angvelnullrot}
\ee
where $b_c=L/E$ is the impact parameter. The double sign in the above equation is related to orbits which are co-rotating and counter-rotating with the BH. Results for co-rotating null orbits are shown in Table \ref{tab:timerotCR} and in Fig. \ref{fig:OmegaCvsJ}. As we expect, results for $J/M^2\rightarrow0$ smoothly tend to the ones for the non-rotating case. Interestingly enough, percentual differences between GR results can be larger than order $5\%$ for co-rotating orbits. The radius for circular null orbits is always smaller for a rotating EDGB BH than for a Kerr BH in GR.
%
\section{Discussion}\label{conclusion}
Einstein-Dilatonic-Gauss-Bonnet theories are viable theories of gravity, which share many features in common with Einstein's gravity
but have a better understood quantum limit. We have improved the linear stability analysis for static black holes in this theory. We found that they are stable and can {\it in principle}
be used to discriminate between the two theories. For the slowly rotating black holes we studied here, the differences in measurable quantities
amount to a few percent in the most favorable cases (large coupling constant). Given the current state-of-the-art, it does not seem possible to distinguish the correct theory of gravity from measurements of either the ISCO (from EM observation) or even of ringdown modes
with future gravitational wave detectors.

There are at least two important extensions to be made, which could prove to be very relevant:

(i) Consider highly spinning black holes. One could attempt to extend the small rotation expansion to second order in $\Omega$, but we found such procedure to be extremely complex, due to the symbolic manipulations required. If carried through, it would allow a computation of other multipole moments of the hole, which could potentially lead to important tests \cite{Berti:2005ys,Will:2007pp}.
Given that many of the astrophysical black hole candidate have a large spin, this could be a promising way of discriminating EDGB and Einstein's theory. To find a highly spinning DBH solution most likely requires a full numerical integration of the field equations (see for instance Ref. \cite{Kleihaus:2007kc} for a concise overview of possible methods to handle this).

(ii) Waveforms of inspiralling particles. The calculation of waveforms is highly non-trivial: it includes either a re-derivation of the PN expansion, or a full-blown numerical solution of the problem. The pay-off is huge: particles orbiting around black holes are able to probe
the background geometry in its entirety \cite{Ryan:1995wh}. The resulting gravitational-waveform {\it should} be sensitive enough to
the field equations, specially at late stages in the inspiral, when the particle is about to merge. Indeed, the particle
spends a large number of cycles near the ISCO, which could potentially increase by orders of magnitude the ability to probe
the geometry better, if compared to ringdown.

Of course much more remains to be done. For instance, it would be desirable to have a quantitative analysis of the lowest lying ringdown modes.
Most of all, it would be extremely important to understand how general theories of gravity affect the strong field regime, the ISCO location and frequency, etc. Perhaps then one might understand how to discriminate between the correct theory of gravity with future electromagnetic or gravitational-wave observations. We hope to tackle these issues in future works.

\section*{Acknowledgements}
We are indebted to Emanuele Berti, Mariano Cadoni and 
Nicolas Yunes for many and useful criticisms and suggestions.
This work was partially funded by Funda\c c\~ao
para a Ci\^encia e Tecnologia (FCT) - Portugal through project
PTDC/FIS/64175/2006. P.P. thanks the CENTRA/IST for hospitality and financial help and the Master and Back fundation programme of Regione Sardegna for a grant. V.C. acknowledges financial support from the Fulbright Scholar programme.

\appendix
\section{Linear stability analysis: axial perturbations}\label{app:linstab}
In this appendix we derive the Schr\"odinger-like equation which represents a eigenvalue problem for the complex frequency $\omega$. Detailed calculations can be found in the online MATHEMATICA notebooks which are available at \verb1http://paolo.casadiale.com/EDGB_BHs.zip1.

The coefficients in eqns. (\ref{eq:32})-(\ref{eq:30}) read 
\begin{widetext}
\beq
A_1(r)&=&\frac{i\,e ^{-\Lambda+{\Gamma}} \left(2  e ^{\Lambda} \left(\Lambda '-{\Gamma} '\right)-\alpha' e ^{{\phi_0}}\left({\Gamma} ' \left(-{\phi_0}' \left(-3 \Lambda '+2 {\phi_0}'+{\Gamma} '\right)-2 {\phi_0}''\right)-2 {\phi_0}'{\Gamma} ''\right)\right)}{2 {\omega}  \left(2  e ^{\Lambda }+\alpha'e ^{{\phi_0}} \left(\Lambda ' {\phi_0}'-2 \left( {\phi_0'}^2+{\phi_0}''\right)\right)\right)}\,,\label{eq:A1}\\
A_2(r)&=&\frac{i\,e ^{-\Lambda+{\Gamma}}\left(-2 e ^{\Lambda}+\alpha' e ^{\phi_0}  {\phi_0}'{\Gamma} '\right)}{{\omega}  \left(2 e ^{\Lambda}+ \alpha' e ^{\phi_0}   \left[\Lambda '{\phi_0}'-2 \left( {{\phi_0}'} ^2+{\phi_0}''\right)\right]\right)}\,,\label{eq:A2}\\
B_{1,l}(r)&=&-\frac{i e ^{-\Lambda}}{2 r {\omega}  \left(e ^{\Lambda } r -\alpha'e ^{\phi_0} {\phi_0}'\right)}\nonumber\\
&&\left\{\alpha'e ^{\phi_0} \left[  -2 e ^{\Gamma} r {\phi_0'} ^2 {\Gamma} '-2 e ^{\Gamma } r {\Gamma} ' {\phi_0}''+{\phi_0}' \left(2 e ^{\Lambda } r {\omega}^2-e ^{\Gamma}\left(\left(e ^{\Lambda }\left(l(l+1)-2\right)-3 r \Lambda '\right) {\Gamma} '+r {{\Gamma} '}^2+2 r {\Gamma} ''\right)\right)\right]  +\right.\nonumber\\
&&+\left. e ^{\Lambda }\left[  -2 e ^{\Lambda }r ^2{\omega} ^2+e ^{{\Gamma}}\left(2 e ^{\Lambda }\left(l(l+1)-2\right)+r \left(r{\phi_0'} ^2+\left(2+r{\Gamma} '\right)({\Gamma} '-\Lambda')+2 r {\Gamma} ''\right)\right)\right]  \right\}\,.\label{eq:B1}
\eeq
\end{widetext}
The explicit for of the $C_{i,l}$ coefficients is not shown here. One can numerically prove that equation (\ref{eq:30}) is automatically satisfied as a consequence of the other two equations and of background solutions. So we are left with a system of two ODEs for the unknown functions $h_0(r)$ and $h_1(r)$. Eliminating $h_0$ from the first order equation (\ref{eq:32}) we obtain a second order differential equation for $h_1$,
\be
A(r)\, h_1''(r)+2 B(r)\, h_1'(r)+C(r)\, h_1(r)=0\,,\label{eq:secord}
\ee
where $A(r)=A_2,\, B(r)=\frac{1}{2}(A_1+A_2'-\frac{2 A_2}{r})$ and $C(r)=A_1'-\frac{2 A_1}{r}-B_{1, l}$.
The function $C(r)$ can be decomposed in $C(r)=Q(r)+\omega^2 E(r)$, where $Q(r)$ and $E(r)$ don't depend on $\omega$. The functions $A, B, E$ and $Q$ depend on the radial background functions $\Gamma, \Gamma', \Gamma'', \Lambda,  \Lambda', \phi_0, \phi_0'$, $\phi_0''$ and on $\phi_0'''$.\\
In order to eliminate the term proportional to $h_1'$ we define
\be
F=\exp{\left(\int_{r_h}^r dr'\;\frac{B(r')}{A(r')}\right)}\,.\label{eq:substF}
\ee
Setting $u=F h_1$, equation (\ref{eq:secord}) takes the Schr\"odinger-like form 
\be
u''(r)+\left[V(r)+\omega^2 K(r)\right]\,u(r)=0\,,
\ee
which is eq. (\ref{eq:oddfinradial}) with
\be
V(r)=\frac{Q}{A}-\left(\frac{B}{A}\right)^2-\frac{d}{dr}\left(\frac{B}{A}\right)\,,\quad K(r)=\frac{E}{A}\,.\label{eq:potoddradial}
\ee
In the limit $r \rightarrow r_h$ the coefficients $A$, $B$, $Q$ and $E$ take the form
\beq
A&=& A_h+{\cal O}\,(r-r_h)\,,\quad B=\frac{ B_h}{(r-r_h)}+{\cal O}\,(1)\,,\\
Q&=&\frac{Q_h}{(r-r_h)^2}+{\cal O}\left( \frac{1}{r-r_h} \right)\,,\\ 
E&=&\frac{E_h}{\gamma_1}\,\frac{1}{(r-r_h)^2}+ {\cal O}\left( \frac{1}{r-r_h} \right)\,.
\eeq
In equations above we have used the following asymptotic behavior near the event horizon
\beq
e^{-\Lambda(r)}&=& \lambda_1 (r-r_h) + ... \,,\\
e^{\Gamma(r)}&=&\gamma_1 (r-r_h) +... \,,\\
\qquad \phi(r)&=&\phi_h+\phi'_h (r-r_h)+...
\label{asymptotia}
\eeq
where
\begin{equation}
\lambda_1=\frac{2}{(\frac{\alpha'}{g^2}e^{\phi_h}\phi_h'+ 2r_h)}~,\label{lambda1}
\end{equation}
and $\gamma_1$ is an arbitrary finite {\it positive} integration constant, which cannot be fixed by 
the equations of motion, since the latter involve only $\Gamma'(r)$ and not $\Gamma (r)$. This constant is fixed by the asymptotic
limit of the solutions at infinity. The asymptotic behavior near the event horizon for $V(r)$ and $K(r)$ is
\beq
V =\frac{V_h}{(r-r_h)^2}\,,\qquad K = \frac{K_h}{(r-r_h)^2}\,,\nn
\eeq
where $V_h\sim1/4$ and $K_h\lesssim4 M^2$. The constants $A_h, B_h, Q_h, E_h, V_h$ and $K_h$ depend on $r_h$, $\phi_h$ and $\phi'_h$. For a Schwarzschild BH, $V_h\equiv1/4$ and $K_h\equiv4 M^2$.
When $r \rightarrow \infty$, $V\rightarrow 0$ and $K\rightarrow 1$. The asymptotic behavior for equation (\ref{eq:oddfinradial}) is described by Eqs. (\ref{eq:oddhor}) and (\ref{eq:oddinf}).

\section{\label{ap:slowGR} Slowly rotating Kerr black holes in the Hartle approximation}
A useful check on the method described in Section \ref{smallrot} is to consider classical GR, and repeat the calculation adding rotation to a Schwarzschild black hole. Perturbing the Schwarzschild metric and retaining only first order terms in $\Omega$, equation (\ref{eq:eqdiffOmega}) reduces to
\be
\Omega_l''(r)+\frac{4}{r}\Omega_l'+\frac{2-l(l+1)}{r(r-2M)}\Omega_l=0\label{eq:eqOmegaS}\,,
\ee
which can be solved analytically in terms of special functions
\be
\Omega_l(r)=C_1\;F(1-l,2+l,4;\frac{r}{2M})+C_2\;G_{20}^{20}\left(\frac{r}{2M}\left|
\begin{array}{c}
-1-l,l \\
-3,0
\end{array}
\right.\right)
\ee
where $F$ is the hypergeometric function and $G$ is the Meijer function. 
The asymptotic behavior of equation (\ref{eq:eqOmegaS}) is the same as eq. (\ref{eq:omasymp}). The asymptotic behavior (\ref{eq:asymOmega}) and the solution at infinity (\ref{eq:omsolasymp}) imply that only the $l=1$ mode is allowed. A list of the general solution of the equation above is:
\beq
&&\Omega_1(r)=C_1+\frac{C_2}{r^3}\,,\quad l=1\,,\\
&&\Omega_2(r)=C_1\left (\frac{r}{M}-2\right )+
C_2{\biggl[}\frac{M^3}{r^3}+\frac{M^2}{r^2}+\frac{3M}{2r}-\frac{3}{2}+\nn\\
&+&\frac{3}{4}\left (\frac{r}{M}-2\right )\log \frac{r}{r-2M} {\biggr]}\,,\quad l=2\,,
\eeq
\beq
&&\Omega_3(r)=C_1\left (\frac{r}{M}-2\right )\left (3\frac{r}{M}-4\right )\nn\\
&+&C_2{\biggl[}4\frac{M^3}{r^3}+10\frac{M^2}{r^2}+30\frac{M}{r}-105+45\frac{r}{M}\nn\\
&-&\frac{45}{2}\left (\frac{r}{M}-\frac{4}{3}\right )\left (\frac{r}{M}-2\right )\log \frac{r}{r-2M} {\biggr]}\,, l=3\,.
\eeq
Demanding a correct asymptotic behavior at infinity, eq.~(\ref{eq:asymOmega}), we verify that $l=1$ is the only mode able to satisfy regularity at the horizon (the result extends to $l>3$ modes, though we do not show those here).
We therefore get the exact result $\Omega=2J/r^3$. This coincides with the expansion (first order in $a=J/M$) of a Kerr metric in Boyer-Lindquist coordinates,
\beq
ds^2&=&\left(1-\frac{2M}{r}\right) dt^2-\left(1-\frac{2M}{r}\right)^{-1} dr^2 \nn\\
&-&r^2\left[d\theta^2+\sin ^2\theta \, (d\varphi-\frac{2aM}{r^3} dt)^2\right]
\label{eq:slowKerr}\,,
\eeq
which we have named Slow Kerr (SK) metric in this work.


\end{document}